\documentclass{aa}  
\usepackage{graphicx}
\usepackage{txfonts}
\usepackage{amsmath}
\usepackage{booktabs} 
\usepackage{dcolumn}  
\usepackage{fixltx2e} 
\usepackage[UKenglish]{isodate}
\cleanlookdateon 
\newcolumntype{d}[1]{D{.}{.}{#1}} 
\def\myeol{\\}

\def\tph{t_{\rm ph}}
\def\dph{d_{\rm ph}}
\def\vph{\mathrm v_{\rm ph}}
\def\peripar{\theta}
\def\nupar{\psi}
\def\tphlin{t_{\rm ph}^{\rm lin}}
\def\dphlin{d_{\rm ph}^{\rm lin}}

\def\tphnom{t_{\rm ph}^{\rm nom}}
\def\dphnom{d_{\rm ph}^{\rm nom}}
\def\vphnom{\mathrm v_{\rm ph}^{\rm nom}}
\def\tphmean{t_{\rm ph}^{\rm av}}
\def\dphmean{d_{\rm ph}^{\rm av}}
\def\vphmean{\mathrm v_{\rm ph}^{\rm av}}
\def\dphhyp{d_{\rm ph}^{\rm hyp}}
\def\vinf{\mathrm v_{\infty}}
\def\ra{\alpha}
\def\dec{\delta}
\def\parallax{\varpi}
\def\pmra{\mu_{\ra\ast}}
\def\pmdec{\mu_\dec}
\def\propm{\mu}
\def\v{\mathrm v}
\def\vr{\mathrm v_r}
\def\vtan{\mathrm v_T}
\def\rsol{r_\odot}
\def\zsol{z_\odot}

\def\deg{$^\circ$} 
\def\arcsec{$''$}
\def\mas{mas}
\def\kms{km\,s$^{-1}$}
\def\maspyr{\mas\,yr$^{-1}$}
\def\Msol{M$_\odot$}

\begin{document}


\title{Close encounters of the stellar kind}
\titlerunning{Close encounters of the stellar kind}
\author{C.A.L.~Bailer-Jones}
\institute{Max Planck Institute for Astronomy, K\"onigstuhl 17, 69117 Heidelberg, Germany}
\date{Submitted 26 October 2014. Revised 26 November 2014. Accepted 11 December 2014.}
\abstract{Stars which pass close to the Sun can perturb the Oort cloud, injecting comets into the inner solar system where they may collide with the Earth. Using van Leeuwen's re-reduction of the Hipparcos data complemented by the original Hipparcos and Tycho-2 catalogues, along with recent radial velocity surveys, I integrate the orbits of over 50\,000 stars through the Galaxy to look for close encounters. The search uses a Monte Carlo simulation over the covariance of the data in order to properly characterize the uncertainties in the times, distances, and speeds of the encounters.  I show that modelling stellar encounters by assuming instead a linear relative motion produces, for many encounters, inaccurate and biased results.  I find 42, 14, and 4 stars which have encounter distances below 2, 1, and 0.5\,pc respectively, although some of these stars have questionable data.  Of the 14 stars coming within 1\,pc, 5 were found by at least one of three previous studies (which found a total of 7 coming within 1\,pc).  The closest encounter appears to be \object{Hip 85605}, a K or M star, which has a 90\% probability of coming between 0.04 and 0.20\,pc between 240 and 470\,kyr from now (90\% Bayesian confidence interval). However, its astrometry may be incorrect, in which case the closest encounter found is the K7 dwarf \object{GL 710}, which has a 90\% probability of coming within 0.10--0.44\,pc in about 1.3\,Myr.  A larger perturbation may have been caused by \object{gamma Microscopii}, a G6 giant with a mass of about 2.5\,\Msol, which came within 0.35--1.34\,pc (90\% confidence interval) around 3.8\,Myr ago.}
\maketitle

\section{Introduction}

Stars passing close to the solar system may threaten life on Earth in at least two ways. First, their gravity can inject Oort cloud comets into the inner solar system where they could impact the Earth \citep{1976BAICz..27...92R, 1982A&A...112..157S, 2002A&A...396..283D, 2011Icar..214..334F, 2014M&PS...49....8R}. Second, hot stars are sources of powerful UV radiation, and supernova remnants produce both $\gamma$-rays and cosmic rays.  If intense enough -- and in particular if the star turned supernova during an encounter -- then such ionizing radiation could kill organisms outright, erode the Earth's ozone layer to expose life to harmful solar UV radiation, or induce long-term global cooling through the NO$_2$ produced in our atmosphere \citep{2005ApJ...634..509T, 2011Ap&SS.336..287B, 2011AsBio..11..551D}.

The geological record shows evidence of extraterrestrial interference. Nearly 200 impact craters are known, many more have long eroded or await discovery, and yet more impacts will have taken place in the oceans \citep{1991Metic..26..175G, EID}.  The mass extinction 65\,Myr ago (between the Cretaceous and Paleogene periods) appears to have been caused, at least in part, by a large impact \citep{2010Sci...327.1214S}, possibly of a long-period comet \citep{2013LPI....44.2431M}.  A layer of $^{60}$Fe on the ocean floor that formed 2.8\,Myr ago is suspected to be the result of a nearby supernova \citep{1999NewA....4..419F, 2002PhRvL..88h1101B, 2004PhRvL..93q1103K}, and the presence of other radioactive isotopes such as $^{129}$I and $^{10}$Be may be indicative of several nearby supernovae over the past few hundred kyr \citep{1996ApJ...470.1227E}. It has been suggested that nearby supernovae could be responsible for mass extinctions \citep{1995PNAS...92..235E}.  For more background on the possible influence of astronomical phenomena on the Earth in general, see \cite{2009IJAsB...8..213B}.

These examples do not mean that stellar encounters represent a significant threat to the terrestrial biosphere, nor that stellar encounters are a major source of Oort cloud perturbation (the Galactic tide also plays a major role here; \citealt{2005A&A...441..783D, 2014MNRAS.442.3653F, 2014M&PS...49....8R}). 
But it is of interest to identify stellar encounters and to examine their association with specific geological signatures.  

The goal of the present work is to identify stars which have passed -- or which will pass -- close to the Sun.  How close a star must come before it has an influence depends on the star's properties.  The magnitude of the perturbation of the Oort cloud can be estimated using the impulse approximation \citep{1976BAICz..27...92R}. For all but the closest encounters this is of order $M/(\vph \dph^2)$, where $M$ is the mass of the star, and $\dph$ and $\vph$ are its distance and speed at closest approach, respectively (the subscript ``ph'' means ``perihelion''). The Oort cloud is believed to extend to 0.5\,pc, but encounters out to a few pc may still produce a significant perturbation if the star is massive and slow.  Concerning ionizing radiation, its intensity will vary as $1/\dph^2$ for a given type of source.  \cite{2011Ap&SS.336..287B} argues that a supernova must occur within 10\,pc to have a significant effect on the biosphere, and OB-type stars would have to come much closer to leave a trace.  A gamma ray burst (GRB) occurring as far away as 1\,kpc could be catastrophic in a similar way to a supernova \citep{2005ApJ...634..509T}. A major source of GRBs is neutron star binaries in globular clusters.  \cite{2013MNRAS.432..258D} have shown that the risk from these has been highly variable over the past 550\,Myr.

In this work I identify individual stars which pass within a few pc of the Sun.  I identify encounters by computing the orbits of stars through a model Galactic potential and computing their distances from the Sun as a function of time. The initial conditions for this numerical integration are the position and velocity derived from astrometric and radial velocity surveys. Such an investigation with a significant number of stars has only been possible since the publication of the Hipparcos astrometric catalogue \citep{1997A&A...323L..49P, 1997ESASP1200.....P} which contains around 118\,000 stars. I analyse here of order 50\,000 of these.  Similar analyses have been performed in the past, most noticeably by \cite{2001A&A...379..634G} who also do a numerical integration, \cite{2010AstL...36..220B} who uses the epicylic approximation to compute orbits, and \cite{2006A&A...449.1233D} who assumed constant stellar velocities (constant potential).  \cite{2001A&A...379..634G} identify 26 encounters with a closest approach less than 2\,pc. These all occur within 4\,Myr of the present time. The closest approach is by \object{Hip 89825} (\object{GL 710}) at 0.38$\pm$0.18\,pc 1.4\,Myr in the future. This is a low mass K7 dwarf, so its impact on the Oort cloud may be minimal.  \cite{2010AstL...36..220B}
and \cite{2006A&A...449.1233D} 
find 13 and 22 stars (respectively) coming within 2\,pc, of which 2 and 7 (respectively) are not in the list of \cite{2001A&A...379..634G} with a perihelion distance below 2\,pc.  These latter two studies also identify \object{Hip 89825} as the closest encounter, albeit at an even closer distance.  There were also some pre-Hipparcos studies (some are listed in the introduction of \citealt{1999AJ....117.1042G}).

Here I perform a new search which offers improvements over previous work.  While some of these improvements were part of some previous studies, no study included all of them.  First, I use astrometry from the new Hipparcos reduction of \cite{2007ASSL..350.....V}, hereafter referred to as Hipparcos-2.  In addition to reportedly improving the accuracy of the astrometry, it has also led to a reduction in the formal errors compared to the original Hipparcos reduction (which was used by \citealt{2001A&A...379..634G} and \citealt{2006A&A...449.1233D}).\footnote{The mean errors in both parallax and proper motion are smaller by about 20\% when measured over the whole catalogue,
  but the improvement is better for the brighter stars (e.g.\ 40\% for V$<$7\,mag).}  Second, I include new radial velocity surveys or re-reductions thereof which increase the number of stars available for analysis.  Third, I adopt a broad initial selection criterion based on unaccelerated relative motion. This is applied to all stars in the catalogues, not just those currently within some arbitrary distance (as done by \citealt{2010AstL...36..220B}), and it avoids assuming some fixed maximum radial velocity (as done by \citealt{2001A&A...379..634G} and \citealt{2006A&A...449.1233D}). Fourth, I use a probabilistic approach to sample over the uncertainties in the astrometry and radial velocities to construct a complete distribution for the perihelion time, distance, and speed for each star. From these I derive a mean and a confidence interval for each parameter.  Using only the nominal values of the data to give a single estimate (as done by most other studies) is biased due to the nonlinear transformations involved \citep{vanAltena2012Brown}.
Armed with the full distributions we can also calculate the (arguably more useful) probability that a star approaches within a certain distance.  \cite{2010AstL...36..220B} also resampled some trajectories to estimate uncertainties.  \cite{2001A&A...379..634G} estimate uncertainties by increasing the parallax, proper motions, and radial velocity individually by 1$\sigma$, calculating new orbits, and then taking the RMS of the four resulting estimates of the perihelion parameters.  Not only is this a very small sample, it is also one-sided (the parameters are all increased), so gives biased uncertainty estimates.

The Hipparcos survey is complete to V\,=\,7.3\,mag \citep{1997A&A...323L..49P}, but 82\% of the stars are fainter than this, and radial velocities are not available for all stars.  I therefore do not claim my search to be ``complete'' in any useful sense of that word. My goal is to search for close encounters, not to perform a statistical analysis on the global nature of encounters.  The search is naturally biased by the available data towards nearby bright stars, and does not include neutron stars (unless they are lurking as undetected companions).  A larger and statistically meaningful analysis will be possible with the upcoming Gaia data \citep{2008IAUS..248..217L}. This should improve the Hipparcos astrometric accuracy by a factor of 50 and extend to G\,=\,20\,mag with a better understood completeness (see section~\ref{sec:future}).

\section{Data}

The basis of this study is the Hipparcos-2 catalogue \citep[CDS catalogue I/311/hip2]{2007A&A...474..653V}. Radial velocities were obtained by cross-matching Hipparcos-2 with three other catalogues: a reanalysis of the Geneva-Copenhagen Survey data (GCS) by \cite{2011A&A...530A.138C} (CDS catalogue J/A+A/530/A138); the Pulkovo catalogue \citep[CDS III/252/table8]{2006AstL...32..759G}; Rave-DR4 \citep[CDS catalogue III/272/ravedr4]{2013AJ....146..134K}. Cross-matching for the first two of these is trivial, as they provide Hipparcos identifiers. Cross-matching with Rave was done using the online CDS cross-matching tool
with a 5\arcsec\ match radius. I also use a fourth catalogue, xhip \citep[CDS catalogue V/137D/XHIP]{2012AstL...38..331A}, which contains both astrometry and radial velocities (for some stars) so no cross-matching was necessary.  This catalogue is based primarily, but not exclusively, on Hipparcos-2.  The reason for this is that Hipparcos-2 gives erroneous solution for a small number of double star solutions which would have required special treatment.
xhip therefore adopts Hipparcos-1 \citep{1997ESASP1200.....P} parallaxes for multiple systems
if the formal parallax error in the Hipparcos-1 solution is smaller. This applies to 1.6\% of the stars. xhip also derives many of its proper motions not from Hipparcos-2, but from Hipparcos-1 and Tycho-2 \citep{2000A&A...355L..27H}, according to the scheme described by \cite{2012AstL...38..331A}.

Whereas GCS and Rave-DR4 give radial velocities from individual surveys, Pulkovo and xhip are compilations of results from various catalogues, possibly the same ones.
Common stars are retained.
Both GCS and Rave-DR4 have some entries with the same Hipparcos ID but different radial velocities. These entries are presumably multiple observations. I retain these as separate entries in my input catalogues. 

The astrometry and radial velocity define the six phase space coordinates of a star in an equatorial coordinate system.  Together with a gravitational potential, these determine the star's orbit (past and present).  I label these coordinates with $D = (\ra, \dec, \parallax, \pmra, \pmdec, \vr)$ and for convenience will refer to all six as the ``astrometric'' data.  Their units -- and that of their measurement uncertainties, denoted with $\sigma$ -- are (deg, deg, \mas, \maspyr, \maspyr, \kms) respectively. ``Year'' throughout this work means the Julian year.

To summarize, this study makes use of four cross-matched input catalogues with the following number of entries in each (only sources with complete astrometric data are retained): hip2gcs (12\,977), hip2pulkovo (35\,483), hip2rave (8727), xhip (46\,368).
Of these 103\,555 entries, there are 51\,251 unique Hipparcos identifiers.
Some of these entries have zero or negative parallaxes (3\% of the Hipparcos-2 catalogue). 
These are excluded from the subsequent analysis.\footnote{In principle they could be retained if the sampling method I use (section~\ref{sec:MethodSampling}) generated a sufficient number of positive parallaxes.} 

I do not purge my input catalogues by updating or removing measurements which appear incorrect, or which have been questioned by other authors (e.g.\ problematic astrometry, possibly due to unaccounted-for multiplicity).  Rather than making ad hoc replacements, I prefer to use well-defined input data sets, and then comment on individual cases in section~\ref{sec:results} (see also the discussion in section~\ref{sec:data_issues}).

\section{Method of perihelion determination}\label{sec:Methods}

\subsection{Overview}

For the purposes of this study, the relevant parameters of the closest approach of a star to the Sun -- its perihelion -- are the time $\tph$, distance $\dph$, and speed $\vph$.  Time is measured with respect to the present day
and distance and speed relative to the solar system barycentre.  I first estimate these parameters for all stars under the assumption of unaccelerated relative motion of the star with respect to the Sun (section \ref{sec:LinearMotion}).  This is quick and easy, but not very accurate, in particular not for stars which are currently far from the Sun or are moving slowly relative to it. So for those stars which have $\dph$ less than some amount (10\,pc), I then compute their orbits and that of the Sun in the gravitational potential of the Galaxy, and calculate new perihelion parameters (section \ref{sec:MethodIntegration}). This is repeated a large number of times for each star by resampling from the covariance matrix of the stellar data in order to construct a frequency distribution over the perihelion parameters (section \ref{sec:MethodSampling}).

\subsection{Linear motion approximation}\label{sec:LinearMotion}

I first neglect gravity and assume that a star moves with constant velocity ${\bf v}$ with respect to the Sun. The position ${\bf r}$ relative to the Sun at time $t$ when the initial ($t=0$) position is ${\bf r}_0$ is
\begin{equation}
{\bf r} \,=\, {\bf r}_0 + {\bf v}t \label{eqn:linearmotion1} \ .
\end{equation}
The perihelion, found from minimization of $|{\bf r}.{\bf r}|$, is at time
\begin{equation}
\tphlin \,=\, - \frac{{\bf r}_0.{\bf v}}{{\bf v}.{\bf v}}
\end{equation}
and at distance $\dphlin$ given by $|{\bf r}|$ from equation~\ref{eqn:linearmotion1}.
We can write these using the measured astrometric data for the initial position and velocity as
\begin{alignat}{2}
  \tphlin \,/\, {\rm yr}\,&=\, -\,c_1&&\frac{1}{\parallax}\frac{\vr}{\v^2} \\
\dphlin \,/\, {\rm pc} \,&=\, 10^3&&\frac{1}{\parallax}\frac{\vtan}{\v} 
\end{alignat}
where
$\parallax$ is the parallax in mas,
$\vr$ is the radial velocity (negative for approaching stars),
\begin{alignat}{2}
\vtan  \,&=\, c_2 \, \frac{\sqrt{\pmra^2 + \pmdec^2}}{\parallax} \hspace{1.5em} &&\text{is the transverse velocity,}\\
\label{eqn:transvel}
\v \,&=\, (\vtan^2 + \vr^2)^{1/2} &&\text{is the total velocity,} \nonumber\\
c_1 \,&=\, 10^3\,{\rm pc}\,{\rm km}^{-1}\,{\rm yr}^{-1} \,&&=\, 0.97779\times 10^9 , \nonumber\\
c_2 \,&=\, {\rm AU}\,{\rm km}^{-1}\,{\rm yr}^{-1} \,&&=\, 4.74047\nonumber,
\end{alignat}
$\pmra$ and $\pmdec$ are the proper motions in \maspyr, and all velocities are in \kms.  Negative values of $\tphlin$ refer to perihelia in the past.

I use the above to calculate the perihelion for all stars and retain only those with $\dphlin<10$\,pc in the subsequent analysis. This limit is chosen to be much larger than distances of interest. I use the nominal values of the astrometric data and do not correct for the bias introduced in $\dphlin$ by the nonlinear transformations in the above equations.  This bias can be neglected for this initial selection because the upper limit on $\dphlin$ is much larger than that of interest.  This is discussed further in sections \ref{sec:completeness} and \ref{sec:perihelion_estimates}.

\subsection{Integration in a Galactic potential}\label{sec:MethodIntegration}

\begin{table}
\begin{center}
\caption{Values of the parameters of the gravitational potential \label{tab:GalaxyParameters}}
\begin{tabular}{lrrr}
\toprule
Component & $M\,/\,M_\odot$ & $a$ / pc & $b$ / pc \\
\midrule
disk & $7.91\times10^{10}$ & 3500 & 250 \\
bulge & $1.40\times10^{10}$ & -- & 350 \\
halo &  $6.98\times10^{11}$ & -- & 24\,000 \\
\bottomrule
\end{tabular}
\end{center}
\end{table}

To improve the accuracy of the perihelion parameters, I integrate the motion of the stars and the Sun through a 3D Galactic potential. I adopt the widely used three-component, time-independent, axisymmetric 
model for the potential from \cite{1975PASJ...27..533M}, which in cylindrical coordinates $(r, \phi, z)$ is
\begin{alignat}{2} 
\Phi \,&=\, \Phi_{\rm d} + \Phi_{\rm g} +\Phi_{\rm h} \\
\Phi_{\rm{d}}   \,&=\, \frac{-GM}{\sqrt{r^2 + \left(a + \sqrt{z^2 + b^2}\right)^2}} \\
\Phi_{\rm{g,h}} \,&=\, \frac{-GM}{\sqrt{r^2 + z^2 + b^2}}
\end{alignat}
where the subscripts $d,g,h$ refer to the Galactic disc, bulge, and halo respectively.
The disk is a highly flattened spheroid with length scales $a$ and $b$. The bulge and halo are spherically symmetric with length scale $b$.  $M$ is the mass of the component. The values of the model parameters are given in Table~\ref{tab:GalaxyParameters}, taken from \cite{1995A&A...300..117D}.  \cite{2001A&A...379..634G} adopted the same potential.  While better models may now exist, the results are not very sensitive to the exact choice. This is because the distances to most stars are not large compared to the scale lengths of the components, so the potentials experienced by the Sun and star are not dramatically different.

It will be shown in section~\ref{ref:potential} that for all encounters found here, the Sun--star interaction is negligible and path deflections by interactions with others stars en route are extremely unlikely, so both may be ignored.

Orbits are computed by numerically integrating the equations of motion through the potential using a seventh order Runge-Kutta-Fehlberg method (rk78f).  The initial conditions for the integration are the six astrometric coordinates of a star, $D$.  These are transformed into Galactic phase space coordinates $(r, \phi, z, \dot{r}, \dot{\phi}, \dot{z})$ using the prescription in \cite{1987AJ.....93..864J}, but updated to the orientation of the Galactic coordinate system in the International Celestial Reference System (ICRS) as given in section 1.5.3 of volume 1 of the Hipparcos catalogue \citep{1997ESASP1200.....P}, which is $(\ra, \dec)_{\rm NGP}$\,=\,(192.85948\deg, 27.12825\deg)
and $\theta_0$\,=\,122.93192\deg.

The coordinate transformation, as well as the computation of the solar orbit, requires us to define the phase space coordinates of the Sun (apart from $\phi$, which is arbitrary in an axisymmetric potential). I adopt $\rsol$\,=\,8\,kpc and $\zsol$\,=\,+10\,pc. The solar velocity is defined in the right-handed UVW coordinate system (vectors parallel to $-r$, $-\phi$, and $z$ respectively). For the U and W components I use the velocity of the Sun with respect to the local standard of rest (LSR) from \cite{2010MNRAS.403.1829S}. These are 11.1\,\kms\ and 7.25\,\kms\ respectively, with uncertainties of the order of 1\,\kms. The LSR is here defined as the point which moves on a circular orbit at $r=\rsol$, $z=0$, and so has zero velocity in the U and W directions. The V velocity of the Sun can be derived more reliably without reference to the LSR by using instead the proper motion in Galactic longitude of the radio source Sagittarius A$^\ast$ at the Galactic Centre. I use a value of $6.379 \pm 0.026$\,\maspyr, equivalently $\dot{\phi}$\,=\,$30.239 \pm 0.123$\,\kms\,kpc$^{-1}$, measured by \cite{2004ApJ...616..872R}.  I assume that this proper motion component is due entirely to the rotation of the Sun about the Galactic Center. Any peculiar motion of the source is likely to be considerably smaller.  With the adopted value of $\rsol$, this corresponds to V$_\odot$\,=\,242\,\kms.

The integration starts at time $t$\,=\,0 (the present) and proceeds to time $t$\,=\,2$\tphlin$ in 1000 steps (the numerical method breaks this down into finer substeps) for both the star and the Sun separately.  Energy and angular momentum were conserved to a very high degree, and finer steps did not change the orbit or perihelia parameters by a significant amount (the numerical imprecision is negligible compared to the uncertainties discussed in section~\ref{sec:discussion}). The step with the minimum separation between star and Sun is taken as the preliminary perihelion (from the linear motion approximation this is expected to occur at the middle step). Although the step sizes are small enough to accurately trace the orbit, they may not always be small enough to determine the perihelion with high accuracy.  The step sizes generally range between 10$^2$ and $10^4$\,yr, during which the separation between the two points around perihelion changes by less than 0.005\,pc in 95\% of cases.
Nonetheless, to overcome the small discretization noise, I apply the linear motion approximation starting from the preliminary perihelion to give the final perihelion parameters. This is simple and physically consistent, yet considerably more efficient than using a much larger number of time steps in the integration.

\subsection{Distribution and confidence intervals from numerical sampling}\label{sec:MethodSampling}

The numerical integration is initially performed using the nominal astrometric data, $D$, to give the perihelion parameters $\tphnom$, $\dphnom$, and $\vphnom$ for time, distance, and speed respectively.  The errors in the data translate into uncertainties in these parameters. Because of the numerical integration, an analytic propagation of these errors is not possible, so I instead sample the error model of the data. The Hipparcos-2 catalogue provides covariance information on the first five astrometric parameters.\footnote{The appendix of \cite{2014A&A...571A..85M} explains how to convert the upper-triangular weight matrix elements provided in the Hipparcos-2 catalogue into a covariance matrix.} I assume the error in the sixth parameter, the radial velocity, to be uncorrelated with the other five. Adopting a 6D Gaussian error model for $D$, I sample it $10^4$ times and use each as the initial conditions of a numerical integration (section~\ref{sec:MethodIntegration}), resulting in $10^4$ sets of perihelion parameters per star. This corresponds to a probability distribution over the parameters. From these I calculate the mean for each parameter, denoted $\tphmean$, $\dphmean$, $\vphmean$.  I choose the mean over the median or the mode because I want a statistic which is influenced by the whole distribution.  I also calculate the 5\% and 95\% quantiles for each parameter to give a 90\% confidence interval. I choose this over the standard deviation because the distributions in $\dph$ are quite asymmetric for the stars of interest, namely those with small $\dph$ (which is strictly non-negative).

Even though the input catalogues only have sources with positive parallaxes, the resampling can generate some samples with negative parallaxes, which are not physical. These samples are rejected.

If the signal-to-noise ratio (SNR) in the radial velocity, $\vr/\sigma(\vr)$, is sufficiently small, then some of the radial velocity samples will have the opposite sign to the nominal radial velocity, resulting in a perihelion time with the opposite sign too. Although this properly expresses the uncertainty in the perihelion time, it renders any single estimate of the perihelion time from this sample (e.g.\ mean or median) problematic.  For this reason I reject samples in which the radial velocity sign differs from the nominal data.

It turns out that sampling from the full covariance matrix produces very similar results to sampling from six one-dimensional Gaussians. I adopt this latter approach for the xhip catalogue, because its astrometry is a variable combination of Hipparcos-1, Hipparcos-2, and Tycho-2, for which the available covariances no longer apply.

I show in the appendix that this data sampling approach is equivalent to estimating the posterior probability distribution over the perihelion parameters given the data, yet is computationally far more efficient. Due to this equivalence, the confidence intervals quoted are Bayesian credible intervals on the estimated parameters.

\section{Results}\label{sec:results}

\subsection{Overall results}

\begin{figure}
\begin{center}
\includegraphics[width=0.5\textwidth, angle=0]{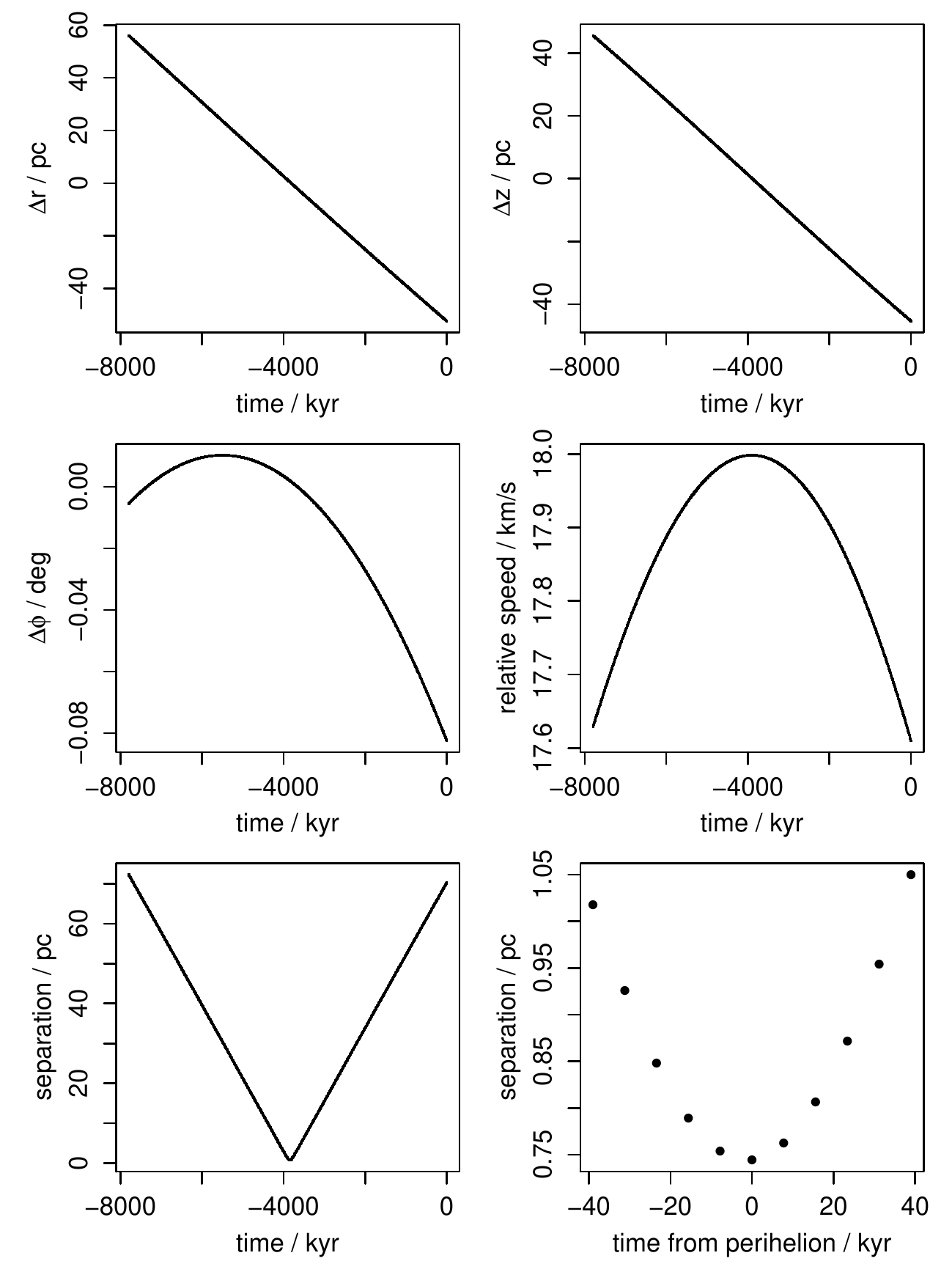}
\caption{The nominal orbit of \object{Hip 103738} (\object{gamma Microscopii}) relative to the Sun in Galactic cylindrical coordinates $(r,z,\phi)$ over the past 7.8\,Myr. The bottom right panel is a zoom of the bottom left panel around the perihelion, and shows the discrete calculation steps in the numerical integration. The perihelion is refined from the closest point using the linear motion approximation. The difference in this case is $3.7\times10^{-4}$\,pc, or 77\,AU.
\label{fig:orbitStar-Sun-2}}
\end{center}
\end{figure}

Some Hipparcos identifiers appear multiple times across the catalogues. I will refer to each unique Hipparcos identifier as a {\em star}, and each catalogue entry as an {\em object}. The latter are indicated using the identifier and a letter to indicate the source catalogue, e.g.\ Hip 12345.x (the letters are defined below).  Applying the linear motion approximation (section \ref{sec:LinearMotion}) to the 103\,555 objects in the four input catalogues, I find 1704 objects (813 stars) with $\dphlin < 10$\,pc.  From now on I will only consider these objects.  The orbits for these (both the nominal data as well as the $10^4$ samples) were then integrated and the perihelion parameters calculated as described in sections \ref{sec:MethodIntegration} and \ref{sec:MethodSampling}. This involved 17 million orbital integrations each taking about 6s on a single CPU core.  An example relative orbit is shown in Figure~\ref{fig:orbitStar-Sun-2}.

The lower the radial velocity SNR, the higher the probability that a star could have a perihelion time with the opposite sign to that reported here.
75 objects have some samples removed due to changes in the radial velocity sign, although of these only 19 have more than 10\% of the samples removed.  Only one of the objects with $\dphmean<2$\,pc, \object{Hip 12351}.x, had any samples removed (0.4\% of them).

The number of objects and stars with mean perihelion distances less than some limit are shown in Table~\ref{tab:dphSummary}.  The data for objects with $\dphmean < 2$\,pc are shown in Table~\ref{tab:periStats} (the online Table at CDS includes all objects out to $\dphmean < 10$\,pc).  This shows the mean values from the resamplings of the three perihelion parameters (time, distance, speed), as well as the 5\% and 95\% quantiles which together form the 90\% confidence interval (CI) on the estimates. The table also shows the fraction of samples which result in $\dph$ less than 0.5, 1.0, and 2.0\,pc.
Negative value of $\tph$ indicate encounters in the past (stars with positive radial velocities).

\begin{table}
\begin{center}
\caption{The number of objects and stars found by the orbit integration to have $\dphmean < \dph^{\rm max}$\label{tab:dphSummary}.
An object is a specific catalogue entry for a star.}
\begin{tabular}{ d{1} r r }
\toprule
\multicolumn{1}{c}{$\dph^{\rm max}$}  & No.\ objects & No.\ stars \\
\midrule
\infty & 1704 & 813 \\
10 & 1548 & 710 \\
 5 &    433 & 213 \\
 3 &    169 &   94 \\
 2 &      65 &   42 \\
 1 &      17 &  14 \\
 0.5 &     5  &  4 \\
\bottomrule
\end{tabular}
\end{center}
\end{table}

\begin{table*}
\centering
\tiny{
\caption{Perihelion parameters for all objects with a mean perihelion distance (mean of the samples; $\dphmean$) below 2\,pc, sorted by this value.
The second column indicates the input catalogue: g=hip2gcs; p=pulkovo; r=hip2rave; x=xhip. Columns 3, 6, 9, are
$\tphmean$, $\dphmean$, $\vphmean$ respectively. The columns labelled 5\% and 95\% are the values of the corresponding confidence intervals.
Columns 12--14 (``\% samples'') indicate the number of samples for each object for which $\dph$ is below 0.5, 1.0, and 2.0\,pc respectively.  Columns 15--20
list the nominal parallax ($\parallax$), proper motion ($\propm$), and radial velocity ($\vr$) along with their standard errors. Potentially problematic objects have not been removed. The online table at CDS includes all objects with $\dphmean<$10\,pc. 
\label{tab:periStats}}
\tabcolsep=0.14cm
\begin{tabular}{*{20}{r}}
\toprule
1 & 2 & 3 & 4 & 5 & 6 & 7 & 8 & 9 & 10 & 11 & 12 & 13 & 14 & 15 & 16 & 17 & 18 & 19 & 20 \\
\midrule
HipID & & \multicolumn{3}{c}{$\tph$ / kyr} &  \multicolumn{3}{c}{$\dph$ / pc} &  \multicolumn{3}{c}{$\vph$ / \kms} & \multicolumn{3}{c}{\% samples} & $\parallax$ & $\sigma(\parallax)$ & $\propm$ & $\sigma(\propm)$ & $\vr$ & $\sigma(\vr)$ \\
          & & av & 5\% & 95\% & av & 5\% & 95\% & av & 5\% & 95\% & 0.5 & 1 & 2 & \multicolumn{2}{c}{\mas} & \multicolumn{2}{c}{\maspyr} & \multicolumn{2}{c}{\kms} \\
\midrule
 85605 & x &    331.64 &    237.01 &    471.38 &   0.10 &   0.04 &   0.20 &    21.0 &    20.5 &    21.5 &  99 &  99 &  99 &   146.8 &    29.8 &     8.6 &     2.8 &   -21.0 &     0.3 \myeol 
 89825 & x &   1387.13 &   1303.13 &   1476.46 &   0.27 &   0.10 &   0.44 &    13.8 &    13.3 &    14.3 &  98 & 100 & 100 &    51.1 &     1.6 &     1.8 &     1.2 &   -13.8 &     0.3 \myeol 
 63721 & p &    146.46 &     89.13 &    240.51 &   0.27 &   0.04 &   0.67 &    34.1 &    27.2 &    41.0 &  90 &  98 &  99 &   216.6 &    56.5 &    41.7 &    58.3 &   -34.0 &     4.2 \myeol 
 89825 & p &   1388.24 &   1302.76 &   1480.74 &   0.36 &   0.12 &   0.64 &    13.8 &    13.3 &    14.3 &  80 & 100 & 100 &    51.1 &     1.6 &     2.3 &     2.1 &   -13.8 &     0.3 \myeol 
 91012 & r &    301.59 &    242.86 &    375.73 &   0.49 &   0.22 &   0.83 &   364.3 &   327.1 &   401.3 &  58 &  98 &  99 &     9.1 &     1.0 &     2.2 &     1.2 &  -364.1 &    22.4 \myeol 
 23311 & r &     10.46 &      9.50 &     11.53 &   0.55 &   0.50 &   0.61 &   815.5 &   736.4 &   894.8 &   5 & 100 & 100 &   114.8 &     0.5 &  1238.2 &     0.7 &  -813.7 &    48.6 \myeol 
 85661 & p &   1877.34 &   1751.16 &   2017.02 &   0.59 &   0.13 &   1.22 &    46.5 &    46.0 &    47.0 &  45 &  87 &  99 &    11.2 &     0.5 &     0.5 &     0.6 &   -46.5 &     0.3 \myeol 
 55606 & r &    133.47 &     94.93 &    188.56 &   0.69 &   0.32 &   1.31 &   919.8 &   772.4 &  1070.1 &  31 &  86 &  98 &     8.3 &     1.5 &     8.9 &     1.6 &  -921.0 &    91.9 \myeol 
 75159 & r &   -343.50 &   -524.47 &   -225.50 &   0.76 &   0.20 &   1.79 &   368.2 &   307.6 &   428.7 &  40 &  82 &  96 &     8.3 &     1.9 &     1.4 &     3.3 &   368.2 &    36.9 \myeol 
103738 & p &  -3851.66 &  -4162.24 &  -3561.42 &   0.83 &   0.35 &   1.34 &    18.0 &    16.7 &    19.3 &  13 &  72 &  99 &    14.2 &     0.3 &     1.8 &     0.3 &    17.6 &     0.8 \myeol 
 71683 & g &     27.71 &     27.65 &     27.77 &   0.90 &   0.89 &   0.92 &    34.2 &    33.9 &    34.6 &   0 & 100 & 100 &   754.8 &     4.1 &  3709.6 &     5.1 &   -25.1 &     0.3 \myeol 
 71683 & p &     27.74 &     27.67 &     27.80 &   0.91 &   0.89 &   0.93 &    34.0 &    33.5 &    34.5 &   0 & 100 & 100 &   754.8 &     4.1 &  3709.6 &     5.1 &   -24.7 &     0.4 \myeol 
 71681 & r &     27.76 &     27.44 &     28.03 &   0.91 &   0.83 &   0.99 &    30.5 &    29.0 &    31.9 &   0 &  96 & 100 &   796.9 &    25.9 &  3702.5 &    28.3 &   -21.0 &     1.0 \myeol 
 70890 & x &     26.71 &     26.64 &     26.76 &   0.94 &   0.92 &   0.96 &    32.6 &    32.0 &    33.2 &   0 & 100 & 100 &   772.3 &     2.4 &  3853.0 &     2.4 &   -22.4 &     0.5 \myeol 
  3829 & x &    -15.07 &    -15.78 &    -14.40 &   0.95 &   0.87 &   1.04 &   269.9 &   262.2 &   277.9 &   0 &  83 & 100 &   234.6 &     5.9 &  2978.2 &     6.1 &   263.0 &     4.9 \myeol 
  3829 & p &    -15.07 &    -15.78 &    -14.38 &   0.95 &   0.87 &   1.04 &   269.9 &   262.1 &   277.8 &   0 &  83 & 100 &   234.6 &     5.9 &  2978.2 &     6.1 &   263.0 &     4.9 \myeol 
 42525 & x &   -251.96 &   -374.97 &   -173.73 &   0.96 &   0.42 &   1.99 &    60.0 &    59.6 &    60.5 &  13 &  68 &  95 &    68.5 &    15.5 &    47.6 &     1.9 &    59.9 &     0.3 \myeol 
 71683 & x &     27.62 &     27.42 &     27.76 &   1.00 &   0.97 &   1.03 &    31.9 &    31.1 &    32.8 &   0 &  48 & 100 &   742.1 &     1.4 &  3709.6 &     2.0 &   -21.4 &     0.8 \myeol 
 57544 & x &     45.05 &     44.44 &     45.67 &   1.05 &   1.02 &   1.08 &   113.9 &   113.7 &   114.1 &   0 &   0 & 100 &   186.9 &     1.7 &   885.2 &     1.9 &  -111.7 &     0.1 \myeol 
 57544 & p &     45.07 &     44.45 &     45.70 &   1.05 &   1.02 &   1.09 &   113.8 &   113.5 &   114.2 &   0 &   0 & 100 &   186.9 &     1.7 &   885.8 &     2.4 &  -111.6 &     0.2 \myeol 
 71681 & x &     26.74 &     25.52 &     27.61 &   1.06 &   1.00 &   1.12 &    30.2 &    28.6 &    31.9 &   0 &   3 & 100 &   742.1 &     1.4 &  3724.1 &    32.7 &   -18.6 &     1.6 \myeol 
 90112 & p &  -1867.53 &  -2110.50 &  -1661.78 &   1.11 &   0.30 &   2.20 &    26.4 &    26.1 &    26.7 &  13 &  48 &  92 &    19.9 &     1.5 &     2.1 &     2.0 &    26.4 &     0.2 \myeol 
 90112 & g &  -1891.92 &  -2138.51 &  -1677.21 &   1.13 &   0.32 &   2.23 &    26.1 &    25.6 &    26.6 &  13 &  46 &  91 &    19.9 &     1.5 &     2.1 &     2.0 &    26.1 &     0.3 \myeol 
 87937 & x &      9.77 &      9.76 &      9.78 &   1.15 &   1.14 &   1.15 &   142.0 &   141.7 &   142.3 &   0 &   0 & 100 &   548.3 &     1.5 & 10308.3 &     2.2 &  -110.5 &     0.1 \myeol 
 87937 & p &      9.73 &      9.72 &      9.74 &   1.15 &   1.14 &   1.16 &   142.3 &   142.0 &   142.7 &   0 &   0 & 100 &   548.3 &     1.5 & 10358.9 &     2.1 &  -110.6 &     0.2 \myeol 
 30344 & x &  -1566.69 &  -1688.43 &  -1454.67 &   1.16 &   0.97 &   1.36 &    18.3 &    17.1 &    19.6 &   0 &   8 & 100 &    34.1 &     0.6 &     5.2 &     0.6 &    18.3 &     0.8 \myeol 
 53911 & r &    -65.16 &    -80.58 &    -52.66 &   1.17 &   0.77 &   1.72 &   819.2 &   738.6 &   900.0 &   0 &  31 &  98 &    18.6 &     2.1 &    67.6 &     2.4 &   818.4 &    48.3 \myeol 
 41312 & r &     47.35 &     44.06 &     50.97 &   1.19 &   1.10 &   1.28 &   681.8 &   631.9 &   730.4 &   0 &   0 & 100 &    30.3 &     0.1 &   156.4 &     0.1 &  -681.2 &    30.0 \myeol 
104644 & r &    -24.52 &    -30.70 &    -19.66 &   1.29 &   0.97 &   1.69 &   599.1 &   479.5 &   721.1 &   0 &   7 &  99 &    67.5 &     3.7 &   720.4 &     3.9 &   596.8 &    73.9 \myeol 
 27288 & x &   -853.93 &   -918.79 &   -793.85 &   1.30 &   1.20 &   1.40 &    24.7 &    23.0 &    26.5 &   0 &   0 & 100 &    46.3 &     0.2 &    14.6 &     0.2 &    24.7 &     1.1 \myeol 
 27288 & p &   -853.82 &   -919.99 &   -794.14 &   1.30 &   1.20 &   1.41 &    24.8 &    22.9 &    26.6 &   0 &   0 & 100 &    46.3 &     0.2 &    14.6 &     0.2 &    24.7 &     1.1 \myeol 
 38965 & x &  -1159.77 &  -1775.51 &   -784.18 &   1.33 &   0.36 &   3.10 &    58.2 &    53.8 &    62.5 &  11 &  46 &  85 &    15.5 &     3.6 &     2.9 &     1.8 &    58.1 &     2.6 \myeol 
 30344 & p &  -1958.39 &  -2048.37 &  -1872.41 &   1.39 &   1.16 &   1.63 &    14.6 &    14.1 &    15.1 &   0 &   0 & 100 &    34.1 &     0.6 &     5.0 &     0.6 &    14.6 &     0.3 \myeol 
 30344 & g &  -1984.84 &  -2060.52 &  -1912.03 &   1.42 &   1.18 &   1.66 &    14.4 &    14.1 &    14.8 &   0 &   0 & 100 &    34.1 &     0.6 &     5.0 &     0.6 &    14.4 &     0.2 \myeol 
 54035 & p &     19.93 &     19.91 &     19.96 &   1.43 &   1.42 &   1.43 &   103.5 &   103.4 &   103.7 &   0 &   0 & 100 &   392.6 &     0.7 &  4801.0 &     0.9 &   -85.8 &     0.1 \myeol 
 54035 & g &     19.95 &     19.88 &     20.02 &   1.43 &   1.42 &   1.44 &   103.4 &   102.7 &   104.1 &   0 &   0 & 100 &   392.6 &     0.7 &  4801.0 &     0.9 &   -85.6 &     0.5 \myeol 
 54035 & x &     20.04 &     20.01 &     20.07 &   1.44 &   1.43 &   1.44 &   102.6 &   102.4 &   102.8 &   0 &   0 & 100 &   392.6 &     0.7 &  4796.6 &     1.7 &   -84.7 &     0.1 \myeol 
 26335 & p &   -490.43 &   -503.82 &   -477.74 &   1.53 &   1.45 &   1.62 &    22.2 &    21.9 &    22.5 &   0 &   0 & 100 &    89.0 &     1.0 &    57.0 &     1.3 &    22.0 &     0.2 \myeol 
 26335 & x &   -489.14 &   -498.78 &   -479.67 &   1.54 &   1.47 &   1.62 &    22.3 &    22.1 &    22.4 &   0 &   0 & 100 &    89.0 &     1.0 &    57.5 &     1.7 &    22.1 &     0.1 \myeol 
 14754 & x &   -289.38 &   -338.36 &   -248.89 &   1.56 &   1.33 &   1.85 &    34.2 &    28.9 &    39.5 &   0 &   0 &  98 &    98.5 &     1.5 &   109.0 &     2.2 &    33.8 &     3.2 \myeol 
 25240 & p &   -991.55 &  -1064.10 &   -924.55 &   1.62 &   1.22 &   2.05 &    55.0 &    54.5 &    55.5 &   0 &   0 &  92 &    17.9 &     0.8 &     6.1 &     0.9 &    55.0 &     0.3 \myeol 
 14754 & p &   -291.24 &   -339.42 &   -250.53 &   1.63 &   1.38 &   1.93 &    34.2 &    29.0 &    39.5 &   0 &   0 &  97 &    97.7 &     1.9 &   111.9 &     3.1 &    33.8 &     3.2 \myeol 
 25240 & g &  -1015.58 &  -1100.24 &   -938.06 &   1.66 &   1.26 &   2.11 &    53.7 &    51.6 &    55.9 &   0 &   0 &  89 &    17.9 &     0.8 &     6.1 &     0.9 &    53.7 &     1.3 \myeol 
 86961 & x &    193.66 &    175.54 &    213.51 &   1.69 &   1.35 &   2.11 &    30.1 &    29.8 &    30.5 &   0 &   0 &  90 &   161.8 &    11.3 &   278.4 &     3.7 &   -29.0 &     0.1 \myeol 
100280 & r &     39.68 &     34.25 &     45.91 &   1.71 &   1.44 &   2.03 &   954.1 &   827.2 &  1084.1 &   0 &   0 &  93 &    26.0 &     0.9 &   226.3 &     1.1 &  -952.9 &    79.2 \myeol 
 30067 & p &   -662.36 &   -678.89 &   -645.90 &   1.76 &   1.64 &   1.87 &    40.6 &    40.4 &    40.7 &   0 &   0 &  99 &    36.3 &     0.6 &    20.2 &     0.8 &    40.5 &     0.1 \myeol 
 30067 & g &   -662.37 &   -679.13 &   -646.30 &   1.76 &   1.64 &   1.87 &    40.6 &    40.4 &    40.7 &   0 &   0 &  99 &    36.3 &     0.6 &    20.2 &     0.8 &    40.5 &     0.1 \myeol 
104256 & r &     57.56 &     50.64 &     65.70 &   1.79 &   1.39 &   2.31 &   919.8 &   894.8 &   944.8 &   0 &   0 &  79 &    18.6 &     1.4 &   118.9 &     1.3 &  -919.1 &    15.1 \myeol 
 30067 & x &   -661.02 &   -677.61 &   -644.88 &   1.81 &   1.70 &   1.93 &    40.7 &    40.5 &    40.8 &   0 &   0 &  99 &    36.3 &     0.6 &    20.9 &     0.7 &    40.6 &     0.1 \myeol 
 86963 & x &    204.41 &    185.48 &    225.41 &   1.82 &   1.45 &   2.27 &    28.4 &    27.9 &    29.0 &   0 &   0 &  78 &   161.8 &    11.3 &   281.1 &    15.7 &   -27.2 &     0.3 \myeol 
 87784 & r &    370.51 &    264.24 &    527.17 &   1.83 &   0.84 &   3.52 &    66.5 &    60.7 &    72.5 &   0 &  11 &  70 &    41.3 &     8.4 &    38.6 &     7.4 &   -66.3 &     3.6 \myeol 
 94512 & x &   3646.96 &   3261.34 &   4076.40 &   1.83 &   0.59 &   3.30 &    30.4 &    29.9 &    30.9 &   3 &  16 &  60 &     8.9 &     0.6 &     0.9 &     0.6 &   -30.4 &     0.3 \myeol 
 12351 & x &   -739.38 &  -1512.31 &   -381.53 &   1.90 &   0.93 &   3.85 &    26.6 &    10.5 &    43.0 &   0 &   8 &  73 &    59.4 &     1.2 &    29.8 &     2.1 &    26.2 &    10.0 \myeol 
 47425 & x &    -63.66 &    -80.29 &    -51.00 &   1.90 &   1.50 &   2.43 &   145.3 &   111.6 &   178.9 &   0 &   0 &  68 &   105.6 &     1.6 &   635.9 &     1.7 &   142.0 &    21.0 \myeol 
 38228 & g &   1294.38 &   1239.32 &   1351.85 &   1.91 &   1.75 &   2.07 &    16.6 &    15.9 &    17.2 &   0 &   0 &  83 &    45.5 &     0.5 &    13.4 &     0.8 &   -16.5 &     0.4 \myeol 
 57548 & x &     71.11 &     70.75 &     71.46 &   1.92 &   1.87 &   1.96 &    37.9 &    37.7 &    38.1 &   0 &   0 &  99 &   298.0 &     2.3 &  1361.2 &     3.0 &   -31.1 &     0.1 \myeol 
 57548 & p &     71.18 &     70.77 &     71.58 &   1.92 &   1.88 &   1.96 &    37.8 &    37.5 &    38.1 &   0 &   0 &  99 &   298.0 &     2.3 &  1361.2 &     3.0 &   -31.0 &     0.2 \myeol 
110893 & p &     88.55 &     87.90 &     89.21 &   1.92 &   1.88 &   1.96 &    38.8 &    38.6 &    38.9 &   0 &   0 &  99 &   249.9 &     1.9 &   980.8 &     3.0 &   -34.0 &     0.1 \myeol 
110893 & x &     88.65 &     88.00 &     89.29 &   1.92 &   1.88 &   1.97 &    38.7 &    38.5 &    38.9 &   0 &   0 &  99 &   249.9 &     1.9 &   980.8 &     3.0 &   -33.9 &     0.1 \myeol 
 26624 & x &  -1871.68 &  -1947.16 &  -1799.57 &   1.94 &   1.74 &   2.14 &    22.2 &    21.6 &    22.9 &   0 &   0 &  69 &    23.5 &     0.4 &     5.0 &     0.3 &    22.2 &     0.4 \myeol 
 38228 & p &   1325.72 &   1297.08 &   1354.90 &   1.95 &   1.81 &   2.10 &    16.2 &    16.0 &    16.3 &   0 &   0 &  70 &    45.5 &     0.5 &    13.4 &     0.8 &   -16.1 &     0.1 \myeol 
 92403 & p &    153.83 &    152.71 &    154.95 &   1.95 &   1.91 &   2.00 &    14.2 &    14.0 &    14.5 &   0 &   0 &  94 &   336.7 &     2.0 &   665.2 &     3.3 &   -10.7 &     0.2 \myeol 
 92403 & x &    153.50 &    152.76 &    154.23 &   1.98 &   1.95 &   2.02 &    14.1 &    14.0 &    14.2 &   0 &   0 &  82 &   336.7 &     2.0 &   668.6 &     2.5 &   -10.5 &     0.1 \myeol 
 34617 & x &  -1224.20 &  -1310.18 &  -1144.86 &   1.98 &   1.45 &   2.56 &    33.6 &    32.7 &    34.6 &   0 &   0 &  53 &    23.8 &     0.9 &     7.7 &     1.7 &    33.6 &     0.6 \myeol 
 26624 & p &  -1871.71 &  -1956.00 &  -1790.91 &   1.98 &   1.73 &   2.24 &    22.2 &    21.6 &    22.9 &   0 &   0 &  55 &    23.5 &     0.4 &     5.1 &     0.3 &    22.2 &     0.4 \myeol 

\bottomrule
\end{tabular}
}
\end{table*}

\begin{figure}
\begin{center}
\includegraphics[width=0.42\textwidth, angle=0]{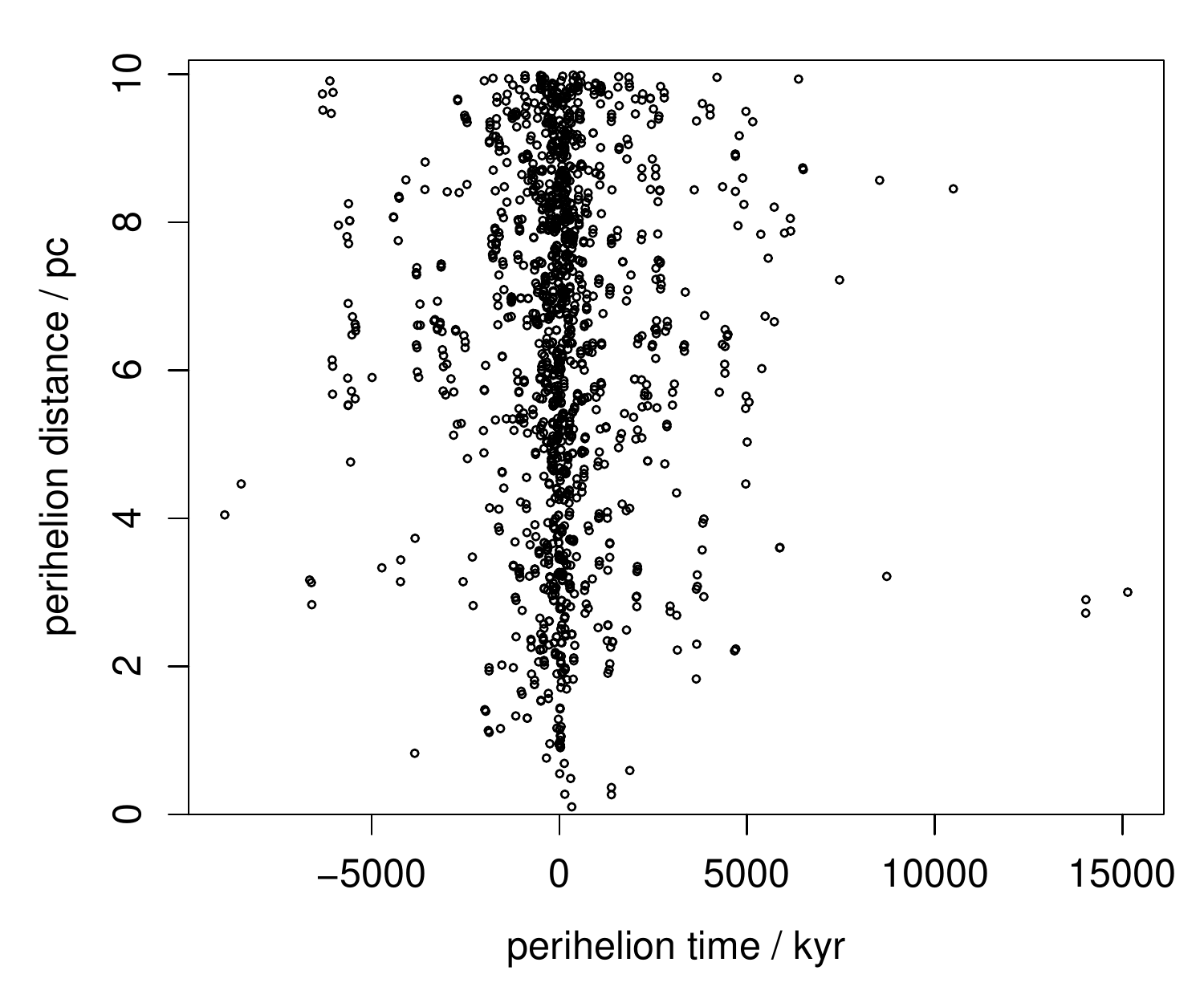}
\includegraphics[width=0.42\textwidth, angle=0]{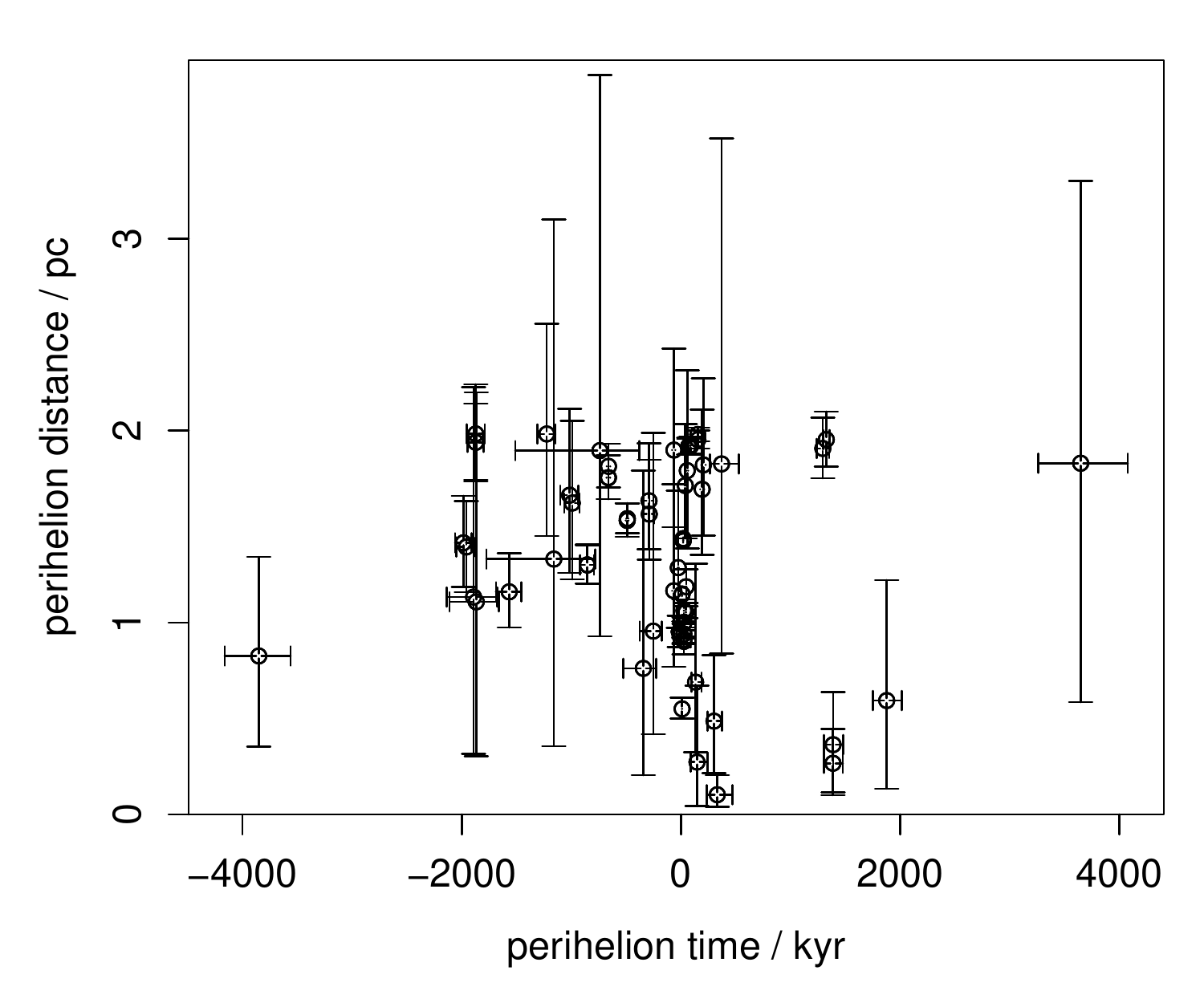} 
\includegraphics[width=0.42\textwidth, angle=0]{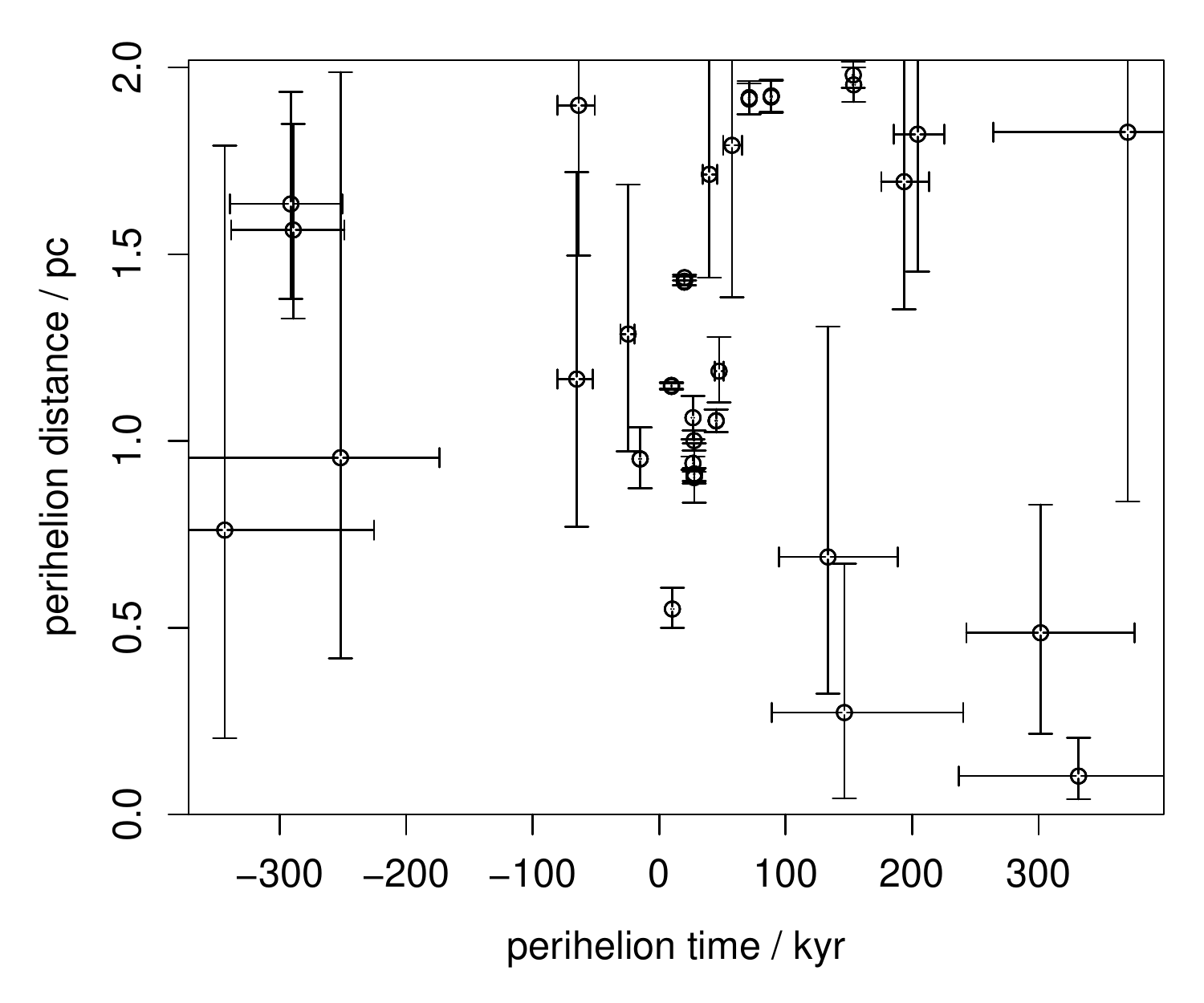} 
\caption{The top panel shows the mean perihelion distance, $\dphmean$, vs.\ mean perihelion time, $\tphmean$, for all stars with $\dphmean<$\,10\,pc. The middle panel shows all stars with $\dphmean<$\,2\,pc and the bottom panel is a zoom of this. The error bars in the lower two panels denote the 5\% and 95\% quantiles of the distributions for each object, which together form a
90\% confidence interval.\label{fig:dph_vs_tph}}
\end{center}
\end{figure}

\begin{figure}[!h]
\begin{center}
\includegraphics[width=0.42\textwidth, angle=0]{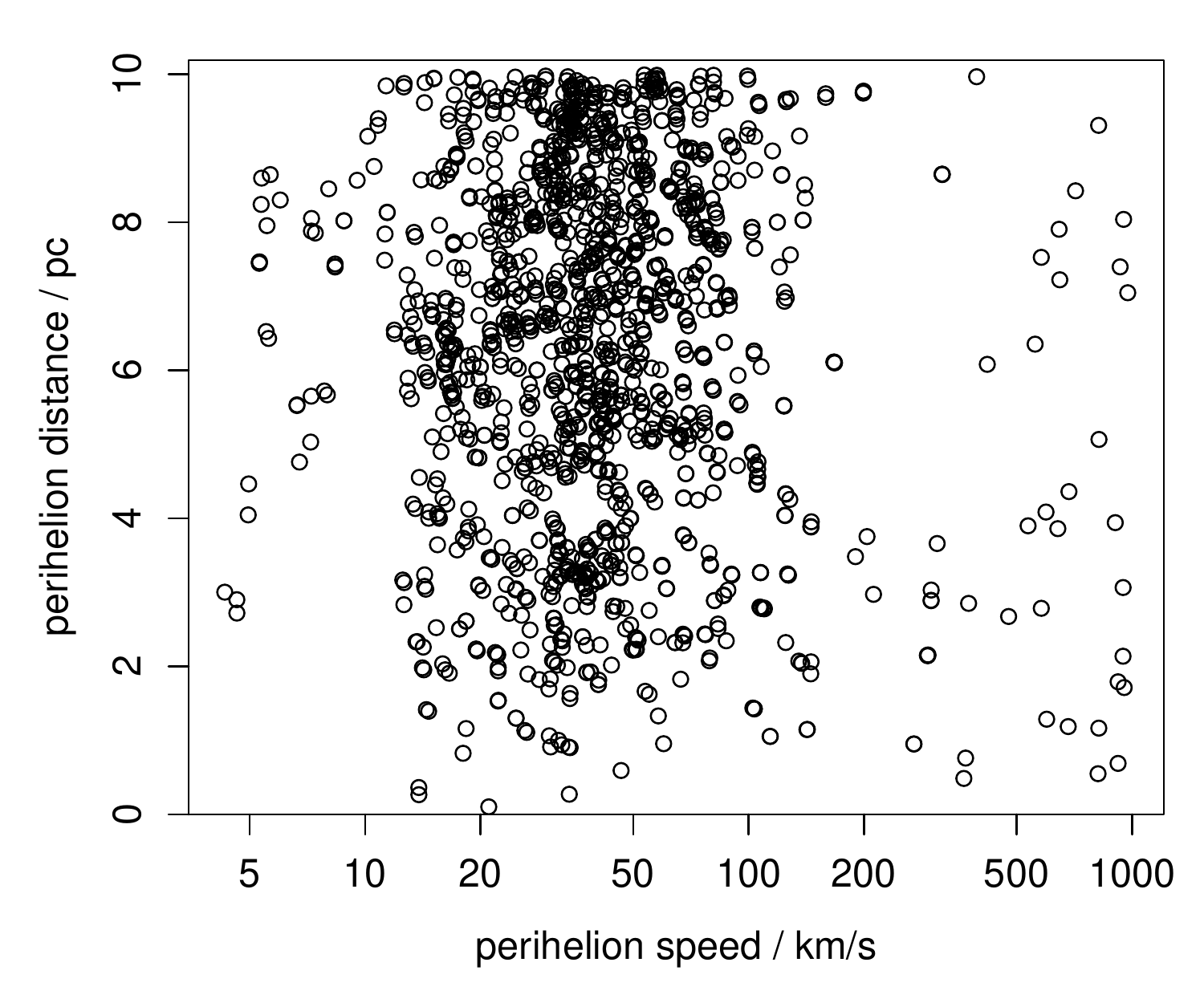}
\caption{The mean perihelion distance, $\dphmean$, vs.\ the mean perihelion speed, $\vphmean$, (on a log scale) for all stars with $\dphmean<$\,10\,pc.\label{fig:dph_vs_log10vph}}
\end{center}
\end{figure}

Fig.~\ref{fig:dph_vs_tph} shows the mean perihelion distances vs.\ time and their confidence intervals. We see a few encounters at 5--15\,Myr in the past or future, but the vast majority occur within 1\,Myr. This is not surprising given the limited magnitude and parallax sensitivity of Hipparcos, both of which correspond to a distance and therefore an encounter time limit. There is an even distribution of past and future encounters: 781 and 766 respectively, which agree within half a standard deviation.
Figure~\ref{fig:dph_vs_log10vph} shows the distribution of the perihelion speeds on a logarithmic scale. We see several objects with very large radial velocities, which I will comment on below.

\begin{figure*}
\begin{center}
\includegraphics[width=1.0\textwidth, angle=0]{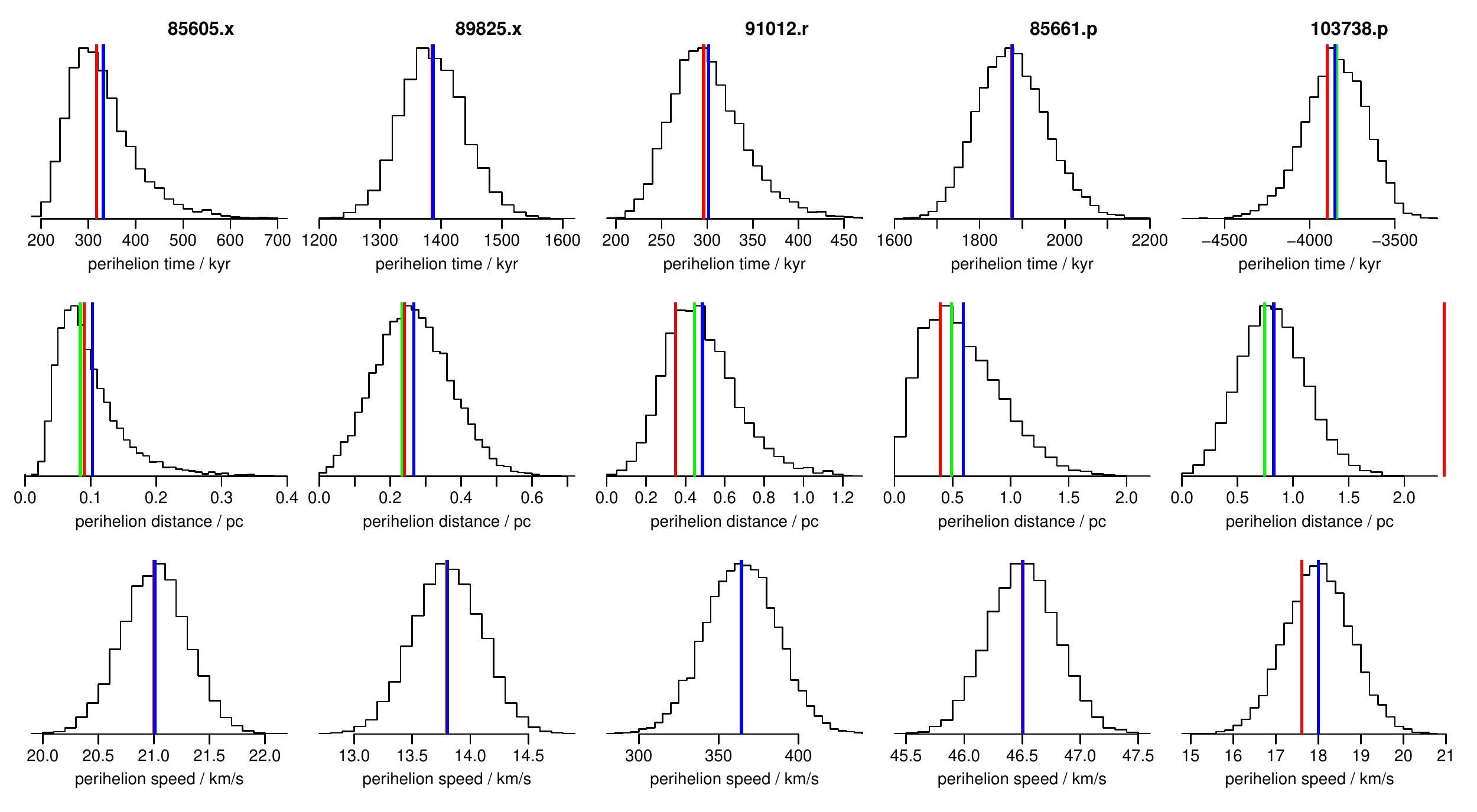}
\caption{Distribution of the perihelion parameters from the resampled data for five selected objects (columns). The rows from top to bottom are
$\tph$, $\dph$, $\vph$. The vertical coloured bars show the three perihelion estimates discussed: the mean of the distribution (blue); the perihelion parameter obtained from using the nominal data (green); the perihelion parameter obtained from the linear approximation method (red). These are sometimes invisible as they coincide with other lines.
\label{fig:perihistogram}}
\end{center}
\end{figure*}

\begin{figure}
\begin{center}
\includegraphics[width=0.45\textwidth, angle=0]{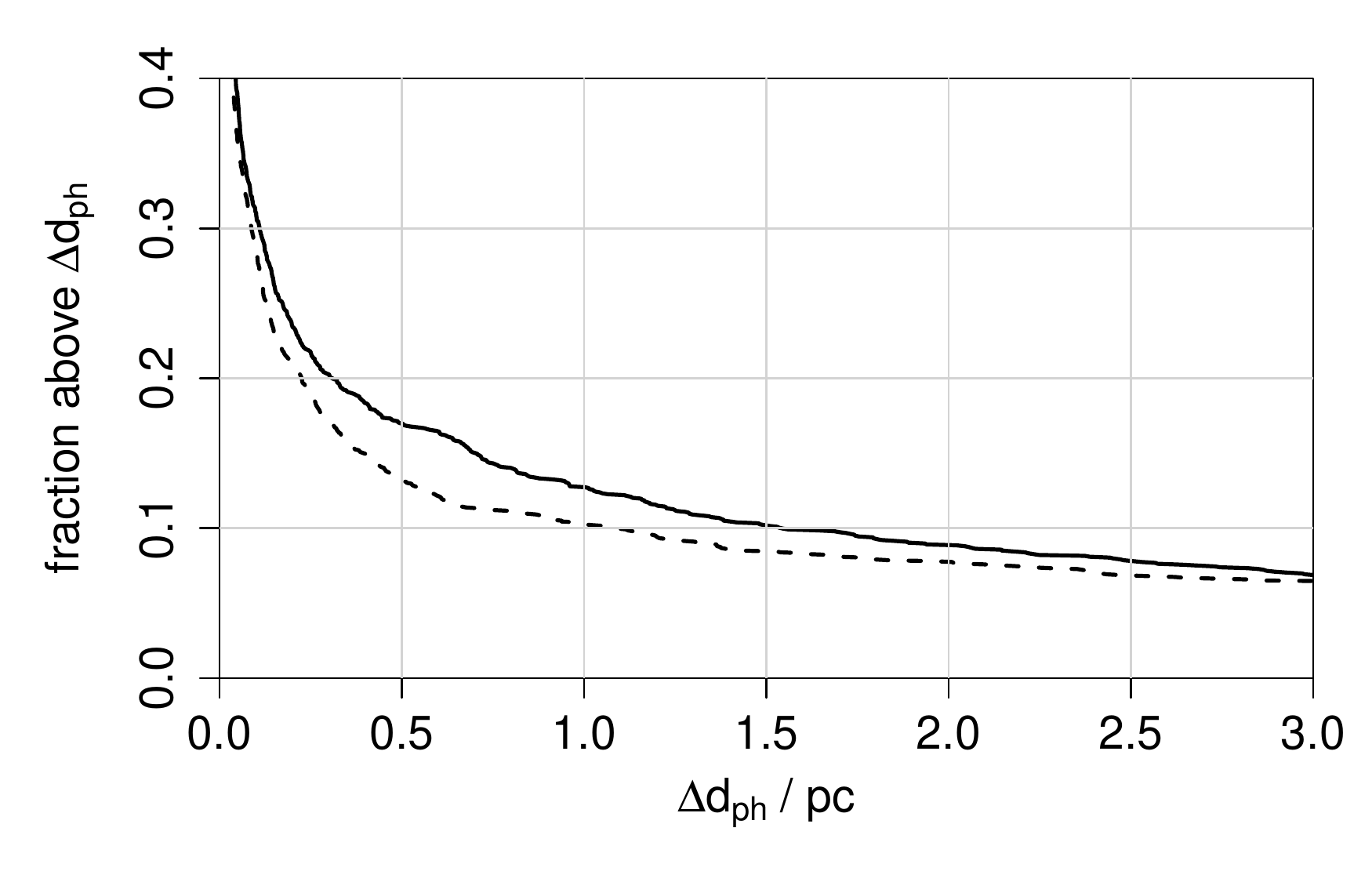}
\caption{Discrepancy between the perihelion distance estimates. The solid line shows the fraction of objects which have a perihelion distance via the linear approximation, $\dphlin$, which differs from $\dphmean$ by more than the amount shown on the horizontal axis, i.e.\ the fraction of objects with $|\dphlin - \dphmean| > \Delta\dph$. The dashed line shows the same but for the nominal distances, i.e.\ 
$|\dphnom - \dphmean| > \Delta\dph$.
\label{fig:dph_discrepancy}}
\end{center}
\end{figure}

The distributions of the perihelion parameters for a few selected objects are shown in Fig.~\ref{fig:perihistogram}.  The asymmetries are apparent (and necessary in $\dph$, as this is strictly non-negative).  Over plotted is the mean of the distributions (blue) as well as the perihelion parameters obtained from the nominal astrometric data in the orbital integration (green) and from the linear approximation method (red).  While the parameters obtained from the three different methods (av, nom, lin) correlate quite well overall,
we see some significant discrepancies in the perihelion distance estimates. Figure~\ref{fig:dph_discrepancy} quantifies this for all objects in the numerical simulation.  It shows the fraction of objects which have perihelion distances ($\dphlin$ and $\dphnom$) differing from the mean perihelion distance by more than some amount. We see, for example, that with the linear approximation method 17\% of objects have distances which differ from $\dphmean$ by 0.5\,pc of more. When calculating $\dph$ using the nominal data in the numerical integration, 13\% deviate by more than 0.5\,pc from $\dphmean$.

The discrepancy in the perihelion times is small, but not insignificant: 17\%, 10\%, and 2\% of objects have perihelion times which differ from $\tphmean$ by more than 25\,kyr, 100\,kyr, and 1000\,kyr respectively (very similar for both $\tphlin$ and $\tphnom$). The agreement for the perihelion speeds is good, with only about 6\% of the objects differing by more than 1\,\kms\ for both methods.

\begin{figure}
\begin{center}
\includegraphics[width=0.45\textwidth, angle=0]{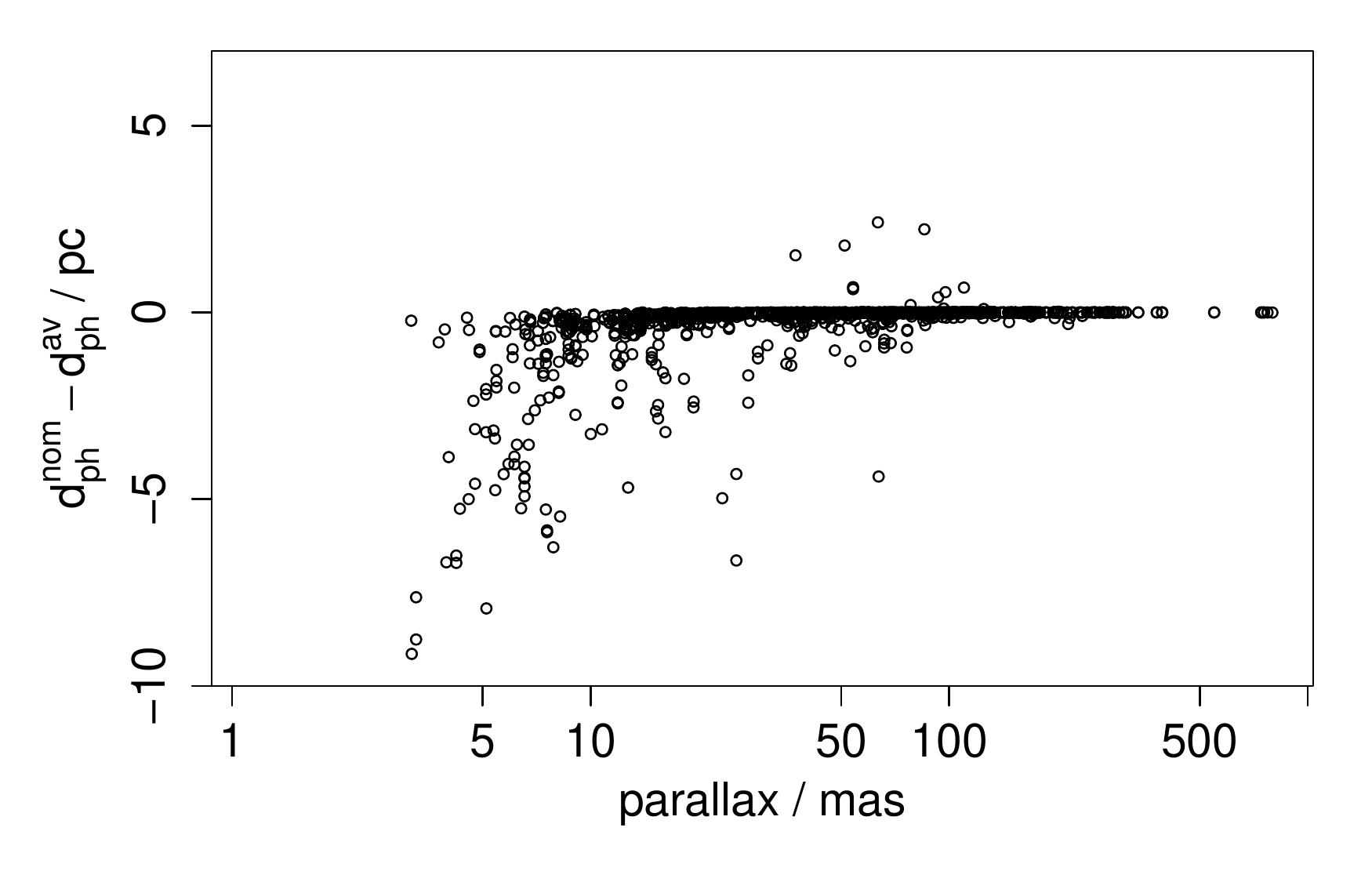}
\includegraphics[width=0.45\textwidth, angle=0]{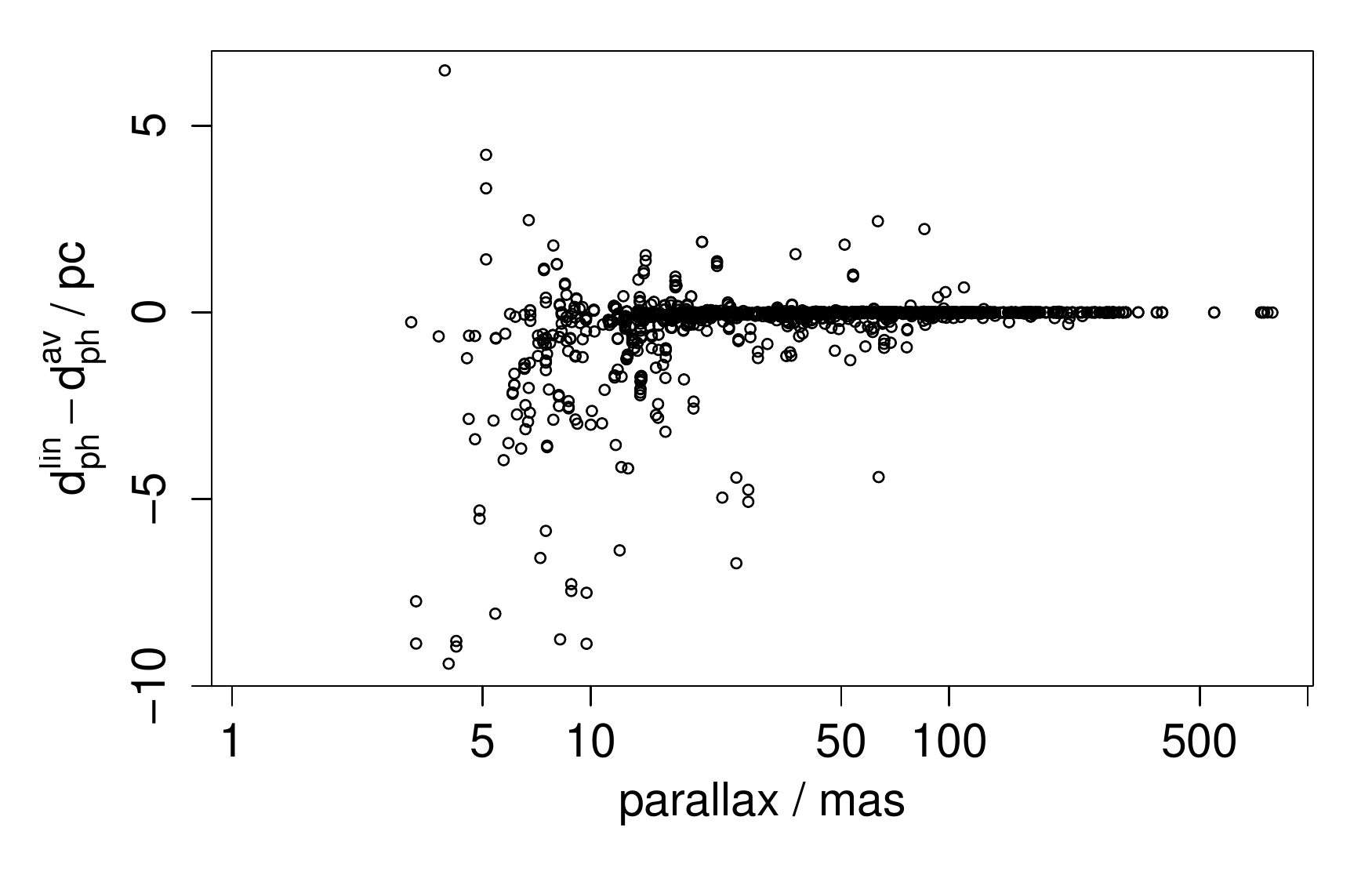}
\caption{Difference between the nominal and mean perihelion distance estimates (top), and linear approximation and mean (bottom), as a function of the parallax on a logarithmic scale. 70 objects with $\dphnom - \dphmean < -10$\,pc 
and 81 objects with $\dphlin - \dphmean < -10$\,pc 
are not shown in the top and bottom panels respectively. A zoom is shown in Figure~\ref{fig:dphdiff_parallax_zoom}.
\label{fig:dphdiff_parallax}}
\end{center}
\end{figure}

\begin{figure}
\begin{center}
\includegraphics[width=0.45\textwidth, angle=0]{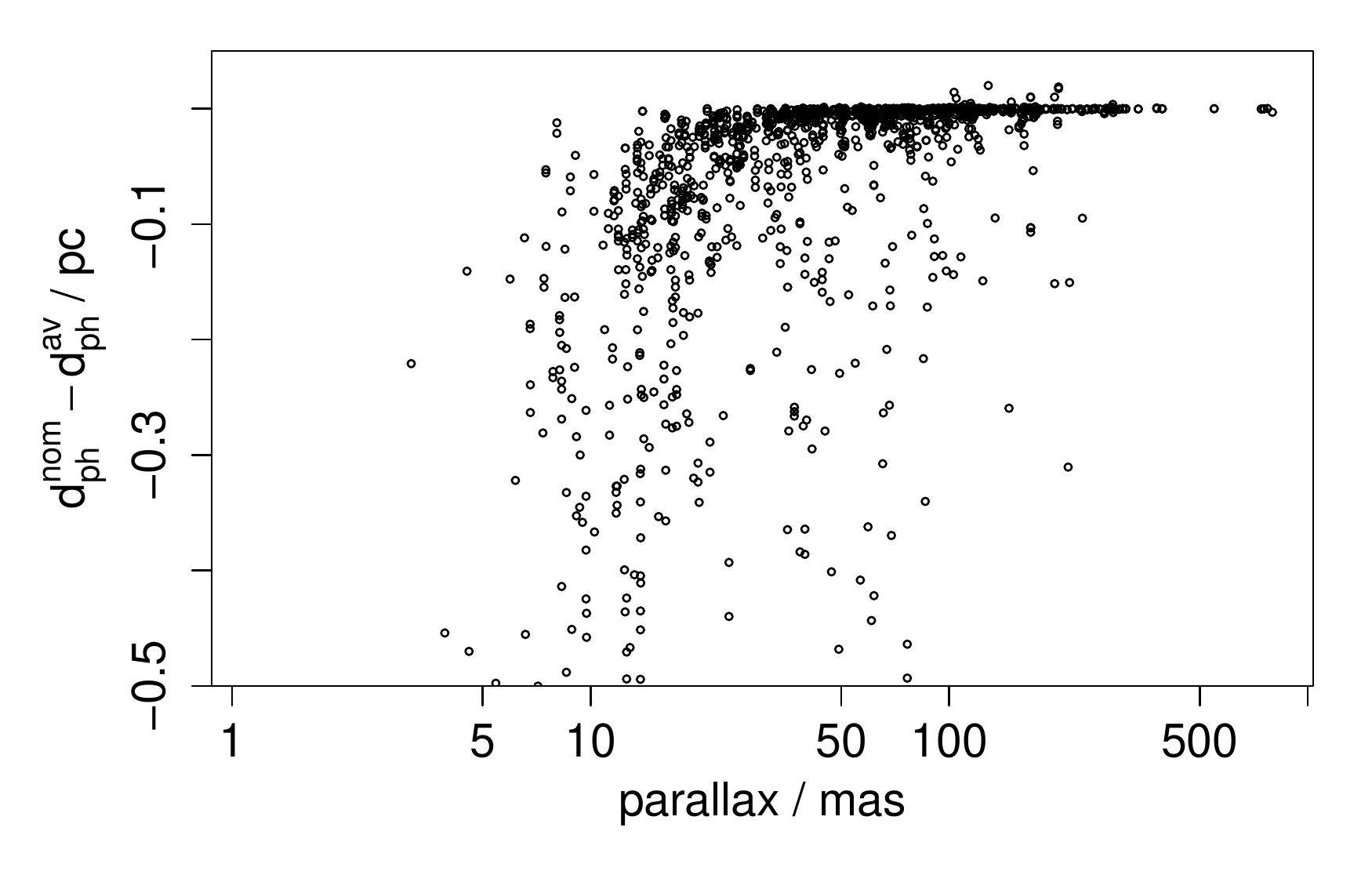}
\includegraphics[width=0.45\textwidth, angle=0]{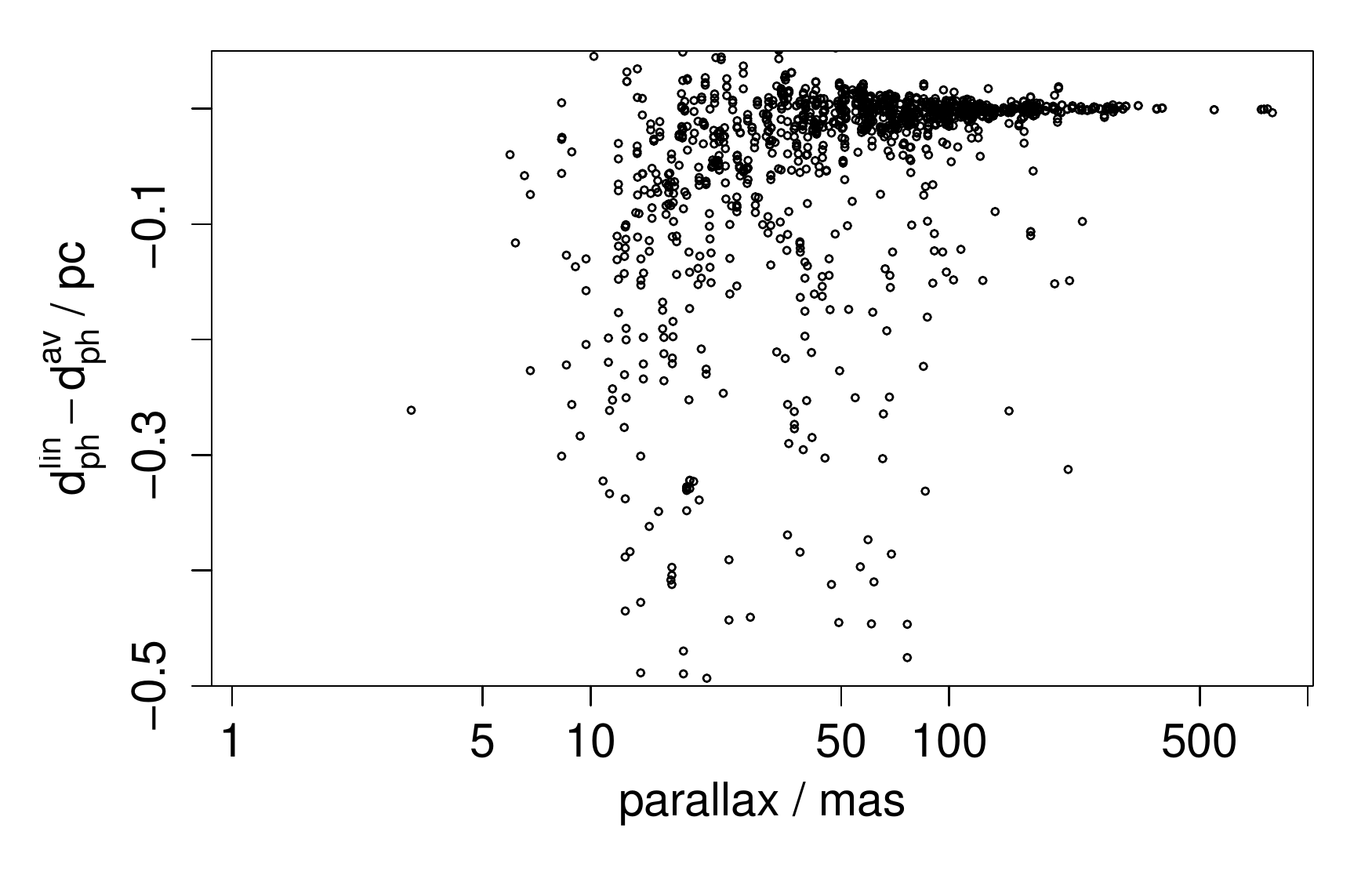}
\caption{As Figure~\ref{fig:dphdiff_parallax} but zooming the vertical axes to highlight the region of smaller differences.
\label{fig:dphdiff_parallax_zoom}}
\end{center}
\end{figure}

We also find that the discrepancies are quite asymmetric in the perihelion distance and speed (less so for the times). The upper panel of Figure~\ref{fig:dphdiff_parallax} shows the discrepancy in the perihelion distance 
from the nominal data as a function of the star's parallax (Figure~\ref{fig:dphdiff_parallax_zoom} shows a zoom of this). We see that using the nominal data tends to underestimate the perihelion distance, and that this gets worse for smaller parallaxes: 1553 objects have distances underestimated, compared to just 151 overestimated, and the underestimations can be quite large. We get similar results when using the linear approximation estimator (lower panel), although less asymmetry (1162 to 542).  Hence both of these other methods will overestimate the number of encounters found and will underestimate the encounter distances. The reasons for this are discussed in section~\ref{sec:perihelion_estimates}.

I find 42 stars which encounter the Sun with $\dphmean<$\,2\,pc (although some of these have problematic data). 25 of these were also found to pass within 2\,pc by at least one of the three previous studies of encounters \citep{2001A&A...379..634G, 2006A&A...449.1233D, 2010AstL...36..220B}. Together, those studies found a total of 34 stars encountering within 2\,pc. If we look only out to 1\,pc, then I find 14 stars compared to their 7, with 5 in common. Put another way, they missed 9 which I found, while I missed 2 identified by them. This difference is usually due to different distance estimates, but in a few cases is due to absent data in one or other of the studies.

\subsection{Individual closest encounters}

The closest encounters are those which could have the biggest impact on the Earth, I examine now those 14 stars which have $\dphmean<1$\,pc.  The letter after the Hipparcos ID below indicates the input catalogue as in Table~\ref{tab:periStats}. Spectral types and colours have been obtained from Simbad (the luminosity class, when absent, is inferred from the absolute magnitude). Those objects marked with $\dagger$ have problematic data, as discussed below. 


$\dagger${\bf \object{Hip 85605}}.x has a 95\% probability of passing within 0.2\,pc, within the Oort cloud.  Its solution ID in Hipparcos-2 is 95, indicating that it is the fainter component of a double system (not necessarily a {\em physical} binary) with a single-star (i.e.\ 5 parameter) astrometric solution. The primary, \object{Hip 85607}, is 22\arcsec\ away.\footnote{The Hipparcos-2 solution ID of 95 for \object{Hip 85607} is an error (F.\ van Leeuwen, private communication).}
This has a negative parallax in Hipparcos-2 and so is excluded from my study.

\object{Hip 85605} also appears in my table as \object{Hip 85605}.p but with $\dphmean$\,=\,3.11\,pc (90\% CI 1.71--5.36\,pc). The difference is due to the xhip catalogue adopting the much smaller Tycho-2 proper motion (on account of the binarity) of $8.6\pm2.8$\,\maspyr, as opposed to $299\pm34$\,\maspyr\ from Hipparcos-2 used in \object{Hip 85605}.p (the parallax and radial velocity are the same).

The Hipparcos catalogue \citep{1997ESASP1200.....P} points out that the astrometric solution for this system is ambiguous.
A note on the Simbad entry for \object{Hip 85605} states that the Hipparcos-1 parallax of 202\,mas and proper motion of 362\,\maspyr\ are erroneous, because the POSS-I and II plates indicate the latter must be less than 50\,\maspyr. This implies that the encounter parameters for \object{Hip 85605}.p -- but not for the xhip entry -- are spurious.  \cite{2000A&AS..144...45F} (see below) give an alternative Hipparcos solution by solving the Hip 85607+85605 system as a physical binary. They derive a parallax and proper motion of $4.6\pm3.7$\,mas
and $10.8\pm4.0$\,\maspyr\ respectively.  These data give
a very large mean perihelion distance of 420\,pc with a huge uncertainty (21--1032\,pc 90\% CI).

F.\ van Leeuwen has kindly reanalysed the Hipparcos astrometry of this system under a number of different assumptions. He concludes that the Hipparcos-2 parallax of \object{Hip 85605} and its (relatively large) uncertainty are valid, but that on account of the large residuals and the complex nature of this system the solution should be treated with caution. For this reason the Tycho-2 proper motion is probably more reliable than the Hipparcos-2 one. If these data are indeed correct, then the very close encounter of \object{Hip 85605}.x is real. Its apparent magnitude of V=11.0\,mag \citep{2000A&A...355L..27H} and parallax of 147\,mas give $M_{\rm V}$=11.8\,mag (assuming a single star with no extinction),
indicating it to be an M dwarf, while its B$-$V colour of 1.1\,mag suggest it has K spectral type.  
Nevertheless, a clear conclusion for this system must await new astrometric data.

{\bf \object{Hip 89825}} (\object{GL 710}), a K7 dwarf, appears twice in the table, from xhip and hip2pulkovo. They differ due to the Tycho-2 proper motion used in the former, although the difference is much less than the uncertainty. This is the closest encounter in the three previous studies, all yielding $\dph$\,=\,0.2--0.4\,pc and $\tph$\,=\,1.4\,Myr. Here I find a probability of 80\% that it passes within 0.5\,pc.

$\dagger${\bf \object{Hip 63721}}.p is solved as a component of double star system in Hipparcos-2.
It appears to be a poor astrometric solution because its large parallax is inconsistent with its spectral type of F3V and apparent magnitude of V=8.7 (from Simbad), which imply it must be much further away. 
\cite{2000A&AS..144...45F} produced new astrometric solutions using both Hipparcos and Tycho-2 astrometry for 257 double and multiple systems which they suspected to have problematic solutions in the Hipparcos catalogue (and they are likely to be problematic in Hipparcos-2 also). They treat this star as a 
component of a binary together with \object{Hip 63716}, and
derive a much smaller parallax of $4.6\pm1.3$\,mas. Their derived proper motion is $40.5\pm1.4$\,\maspyr, similar in size to, but with a much smaller uncertainty than, the Hipparcos-2 value. Thus the encounter listed in Table~\ref{tab:periStats} should be treated as dubious.

$\dagger${\bf \object{Hip 91012}}.r is listed by Simbad as an F5 star and has an effective temperature from Hipparcos/2MASS data \citep{2011MNRAS.411..435B} of 6650\,K (90\% CI 6350--6910\,K). It is therefore likely to be slightly more massive than the nearer encounters discovered, so potentially has a larger perturbing effect on the Oort cloud. However, the perihelion speed is large, 360\,\kms, which reduces the impulse transfer. This large speed is a direct result of the large radial velocity in Rave of $-364\pm22$\,\kms, which seems suspicious (see section~\ref{sec:data_issues}). This star appears another three times in my list: another measurement from hip2rave ($\vr=-36.5\pm18.7$\,\kms) and one from each of hip2gcs and hip2pulkovo (both $\vr=-16.8\pm0.5$\,\kms).  The latter two both give $\dphmean=8.6$\,pc with a combined 90\% CI of 3.6--15.4\,pc.

$\dagger${\bf \object{Hip 23311}}.r (\object{GJ 183}) is a high proper motion K3 dwarf and one of the closest encounters in terms of time. This star also appears in the other three input catalogues with the same astrometry, but with much smaller radial velocities: around $21\pm0.3$\,\kms\ as opposed to $-814\pm49$\,\kms\ in Rave (cf.\ a transverse velocity from equation~\ref{eqn:transvel} of 51\,\kms).
Consequently those other objects have more distant encounters of about 8\,pc around 60\,kyr ago. The Rave radial velocity is so high that, if correct, this star would not be bound to the Galaxy as described by the potential used (but see section~\ref{sec:data_issues} for a possible problem with these large Rave radial velocities).

{\bf \object{Hip 85661}}.p is an F dwarf in a binary system. It also appears in the xhip catalogue with a ten times larger nominal proper motion from Tycho-2. This results in a larger perihelion distance of 2.75--5.61\,pc (90\% CI). This star is the second-closest encounter found by \cite{2001A&A...379..634G} at $\dph=0.94\pm0.71$\,pc.

$\dagger${\bf \object{Hip 55606}}.r is a G0 dwarf. Its solution ID in the Hipparcos-2 catalogue is 1, indicating that it is a stochastic solution taken from the Hipparcos-1 catalogue. Solutions are classified as stochastic when their residuals are larger than expected, which probably indicates the presence of unresolved orbital motion in a multiple system \citep{2007ASSL..350.....V}.
It also has a very large radial velocity from Rave. Its corresponding Galactocentric space velocity of 857\,\kms\ is considerably higher than the escape velocity of 574\,\kms.

$\dagger${\bf \object{Hip 75159}}.r is an F5 dwarf with another large radial velocity in Rave.

{\bf \object{Hip 103738}}.p (\object{gamma Microscopii}) is a G6 giant which encountered the Sun about 3.9\,Myr ago.
With a mass of 2.5\,\Msol, this is probably the most massive encounter with $\dphmean<$\,1\,pc.
The xhip entry for this star (\object{Hip 103738}.x) uses the Tycho-2 proper motion which is about twice as large in magnitude, resulting in a more distant encounter ($\dphmean=3.7$\,pc, 90-\% CI 2.3--5.2\,pc) but at the same time.  \cite{2001A&A...379..634G} found this object to encounter at $\tph=-3750\pm240$\,kyr and $\dph=1.25\pm1.01$\,pc, which is consistent with the closer of my entries.

{\bf \object{Hip 71683}} (\object{alpha Centauri A}) is of course a well studied star, and appears in three of my four input catalogues (not hip2rave).  One of the closest stars to the Sun at 1.32\,pc, it will move to its perihelion of 0.90\,pc in 27\,kyr.  The xhip catalogue adopts a smaller parallax (and a smaller proper motion error) resulting in a slightly more distant encounter.  

{\bf \object{Hip 71681}} (\object{alpha Centauri B}) is the binary companion of alpha Cen A.

{\bf \object{Hip 70890}} (\object{proxima Centauri}) is probably the third component in the alpha Cen system and is the closest known star to the Sun.

{\bf \object{Hip 3829}} (\object{van Maanen's star}) is the third closest white dwarf to the Sun and the closest known solitary white dwarf. It appears in hip2pulkovo and xhip with identical data. (The resulting perihelion parameters are not identical, however, because of the independent random sampling involved in deriving them.)  It encountered the Sun just 15\,kyr ago.  \cite{2001A&A...379..634G} found a much larger perihelion distance of $3.33\pm0.14$\,pc, probably because they used a smaller radial velocity of $54\pm3$\,\kms\ (the Hipparcos-1 and Hipparcos-2 astrometry, in contrast, are very similar).

$\dagger${\bf \object{Hip 42525}}.x has a solution ID in the Hipparcos-2 catalogue of 1 (stochastic solution).  It has a close companion which surely effects some measurements, as its spectral type is F8 (consistent with its colour of $B-V$\,=\,0.6), yet its nominal absolute magnitude is $M_V$\,=\,9.5, which is typical of an early M dwarf.
The general notes of the Hipparcos-1 catalogue lists a ``more likely'' solution with a very different (and barely significant) parallax of $5.08\pm4.28$\,\mas. 

\subsection{Massive stars}

While the perihelion distance is a major factor in the possible effect of a star on the solar system, the mass and encounter speed also play a role. The impulse gained by an Oort cloud comet from an encounter is of order $M/(\vph \dph^2)$ \citep{1976BAICz..27...92R}.  Masses can be crudely estimated from the spectral type or colours, but I do not investigate the impulse systematically in this work. I note, however, that there are 17 B and A-type stars which have encounters with $\dphmean<4$\,pc from at least one of the input catalogues, including \object{Sirius} (\object{Hip 32349}) at 2.51\,pc in 46\,kyr and \object{Altair} (\object{Hip 97649}) at 2.65\,pc in 138\,kyr. The closest approach is by \object{zeta Leporis} (\object{Hip 27288}), an A2 dwarf or subgiant which attained $\dphmean$\,=\,1.3\,pc about 850\,kyr ago.  An interesting case is \object{Algol} (\object{Hip 14576}), which is a triple system dominated by a B8 dwarf with a total mass of around 6\,\Msol.  With $\dphmean$\,=\,4.1\,pc (90\% CI 0.83--15.4\,pc) from xhip (hip2pulkovo gives similar values), its perturbing effect would not be more than some closer, less massive encounters, despite the low encounter speed ($\vphmean$\,=\,5.0\,\kms, 90\% CI 0.8--10.5\,\kms). However, its radial velocity is close to zero ($3.9\pm3.7$\,\kms), so it is not possible to know whether the encounter is in the past or future. The estimate here of $\tphmean=-8910$\,kyr, (90\% CI $-$25.8 to $-$2.6\,Myr) is from adopting just positive radial velocities in the sampling.

There are also another 12 giant stars (in addition to \object{gamma Microscopii}) which passed (or will pass) within $\dphmean<4$\,pc of the Sun,
as well as one supergiant, \object{Hip 87345} (\object{RY Sco}), a classical Cepheid with $\dphmean$\,=\,3.6\,pc albeit with a wide confidence interval (0.9--7.6\,pc).

\section{Discussion}\label{sec:discussion}

\subsection{Completeness}\label{sec:completeness}

This study makes no claim to be a complete survey for encounters out to some distance or magnitude limit, because it is limited by the input catalogues selected. But we can consider how efficiently it has detected encounters from the given data.  An incompleteness arises because not all 103\,555 objects were subject to numerical orbit integration, due to computational time constraints.
Instead, the linear approximation method was used initially to select encounters with $\dphlin<10$\,pc and only those integrated.  Given that this approximation is not perfect (see Figure~\ref{fig:dph_discrepancy}), there are presumably some objects with $\dphlin>10$\,pc which would have achieved $\dphmean<10$\,pc had they been selected.  Given this initial selection limit of $\dphlin<10$\,pc, then the larger the upper bound of $\dphmean$, the worse the completeness of the sample.  We can estimate the magnitude of this as follows: 297 objects have $9<\dphlin<10$\,pc, so would have been missed had I made an initial selection using $\dphlin<9$\,pc. Of these, only 17 (6\%) end up with $\dphmean<9$\,pc.
The fraction missed will be lower for smaller values of $\dphmean$.  This is not an accurate measure of completeness, but it does suggest that the fraction of close encounters overlooked
is very small.

\subsection{Differences between the perihelion estimates}\label{sec:perihelion_estimates}

We saw significant discrepancies between the perihelion estimates obtained using just the nominal data and those obtained using the full set of resampled data (e.g.\ Figure~\ref{fig:dphdiff_parallax}).  In particular, the perihelion distance from the former was preferentially underestimated. Part of the reason for this is the nonlinear transformation between parallax and distance. A symmetric distribution in the parallax uncertainty corresponds to an asymmetric distribution in the distance: If the mean parallax is $\parallax$\,=\,15\,mas and a fraction $f$ of the probability lies in the range 5--15\,mas and thus also in 15--25\,mas, then once transformed to distance, the same fraction $f$ lies in 200--67\,pc and in 67--40\,pc.  Thus the mean of the distance distribution is larger than the nominal distance, $1/\parallax$.  Likewise, the mean of the perihelion distances from the resampled data will generally be larger than the nominal one, explaining why $\dphnom$ tends to be underestimated. The bias is larger for smaller parallax SNRs, which is what we see in Figure~\ref{fig:dphdiff_parallax} (parallax SNR is correlated with parallax in this Hipparcos sample).

The linear approximation method also suffers from this bias, and additionally from the fact that it neglects gravity.  Nonetheless, it is interesting to observe that the speed is not much changed by gravity. For the 1704 objects with $\dphlin<$\,10\,pc, only 6\% experience a speed change of more than 1\,\kms\ between the current time and their perihelion (and only 1\% more than 10\,\kms).  But gravity does change the direction of motion, and this combined with the parallax bias results in poorer estimates of $\tph$ and $\dph$ from the linear approximation method.

Given the magnitude of the discrepancies found in section~\ref{sec:results}, it is clear that we should not use the linear approximation or the nominal data alone to identify close encounters.

\subsection{Dependence on the gravitational potential}\label{ref:potential}

The perihelion parameters found by the orbital integration of course depend on the Galactic potential adopted. I intentionally chose a simple three-component axisymmetric potential without bar or spiral arms.\footnote{Any asymmetry is poorly constrained by currently available data; introducing a tentative model will not make the results more robust.}
What is relevant for the orbits is the difference in the potential which the Sun and the star experience, and this will remain small provided the separation between them remains small compared to the length scales of the potential (see Table~\ref{tab:GalaxyParameters}). The nominal distances ($1/\parallax$) of the 1704 stars selected by the linear approximation extend up to nearly 900\,pc, so there is sensitivity, although for stars with $\dphmean$\,<\,10\,pc and <\,2\,pc the maximum nominal distance is 315\,pc and 120\,pc respectively.  I also chose not to sample over the current phase space coordinates for the Sun. Provided the uncertainty in the position of the Sun is small compared to the length scales of the potential, which is surely the case, this will not be a significant source of error.

\cite{2001A&A...379..634G} investigated the sensitivity of their results to changes in the potential. (I use the same potential model and parameters, same values of $\rsol$ and $\zsol$, and  a similar solar velocity.)
Varying $\zsol$ from 0--20\,pc and $\rsol$ from 7.5--8.5\,kpc had negligible influence on the calculated perihelion distances.  Changing the local mass density by a factor of two resulted in only 8 of their 156 stars with $\dph$\,<\,5\,pc having their perihelion distances changed by more than 0.05\,pc (up to 2.2\,pc).
They also add a spiral arm potential to their model and find that only 8 of the 156 stars have their perihelion distances changed by more than 0.1\,pc (their Figure 10a).

In common with previous studies, I ignore the gravitational attraction between the Sun and encountering star. In reality they move on hyperbolic orbits with respect to one another, and the true perihelion will be smaller than calculated, 
the discrepancy increasing with smaller $\vph$ and $\dph$ and with larger stellar mass. 
From the theory of orbits (e.g.\ \citealt{Curtis}), the perihelion distance when only this force is acting is
\begin{equation}
\dphhyp \,=\, \sqrt{a^2 + d^2} - a
\end{equation}
where $a = GM_t/\vinf^2$ (the semi-major axis), $M_t$ is the sum of the masses, $d$ is the impact parameter (the perihelion distance without any gravity), and $\vinf$ the speed of the star (relative to the Sun) at infinity. 
Yet even for the most deviant case imaginable among my close encounters -- $\vinf$\,=\,10\,\kms, $d$\,=\,0.05\,pc, $M_t$\,=\,6\,\Msol\ -- 
the perihelion distance is reduced by only 0.5\% when taking into account the Sun--star interaction.

My potential is smooth and time-independent. The smoothness means that interactions of the Sun (or encountering stars) with individual stars or molecular clouds along their orbit are neglected, yet in principle these could deflect the path from the orbit I have calculated. The encounters found in this work take place between -16.5 and +19.6\,Myr from now (the range of $\tphmean$ for all encounters).
Over the course of 20\,Myr the Sun moves about 330\,pc with respect to the LSR and thus typical stars in its neighbourhood.
Let us consider an interaction as significant if it results in a deflection of the Sun's path of 0.5\deg\ (or more). 
Considering this interaction again as a hyperbolic orbit, the deflection angle is
\begin{equation}
\alpha \,=\, 2\arctan\left(\frac{a}{d}\right) \,\simeq\, \frac{2a}{d} 
\end{equation}
the approximation being relevant for small angles.
We get a 0.5\deg\ deflection when the Sun encounters a 1\,\Msol\ star at 34\,\kms\ (the median perihelion speed of the encounters in this work) 
with an impact parameter $d$\,=\,0.0017\,pc.
As an approximation, I consider an interaction to take place only if the Sun approaches another star within this distance. Assuming a stellar number density of $\rho$\,=\,1\,pc$^{-3}$ (which is quite high), then the average number of encounters after travelling a distance $x$ is $\lambda$\,=\,$\pi d^2 x \rho$. The probability of $n$ encounters is given by the Poisson distribution with rate parameter $\lambda$. The probability of no encounters after $x$\,=\,330\,pc is therefore $e^{-\lambda}$\,=\,0.997.
All of the encounters modelled in this work occur within a shorter path length than this, so neglecting the discreteness of the gravitational potential is justified in this study. A similar argument can be applied to the encountering stars.

A time dependency of the potential could arise from its intrinsic evolution or from the Sun experiencing small scale changes in the potential as it moves.  The Sun will move about 30\deg\ around the centre of the Galaxy in 20\,Myr (the period of the LSR in my model is 218\,Myr), although of course the Galaxy itself is rotating too.  The impact of this would have to be modelled with specific asymmetries, but given that their inclusion was found by \cite{2001A&A...379..634G} to have a small impact, so would their variation.

\subsection{Issues with the data}\label{sec:data_issues}

Aside from the method and the model assumptions, we should also scrutinize the reliability of our data and the estimated covariances.  As discussed in section~\ref{sec:results}, some of the astrometric solutions for double or multiple systems may be erroneous. The issues are discussed further in volume 1 of the Hipparcos catalogue \citep{1997ESASP1200.....P}.
Some alternative solutions have been mentioned, but ultimately new data may be required. I have assumed that the measured radial velocities reflect the space motion of the star relative to the solar system barycentre, whereas some may be affected by motion in a binary system. More radial velocity epochs would be required to identify this.

The Rave-DR4 catalogue contains a significant number of objects with much larger absolute radial velocities than the other input catalogues. The central 50\% quantile of $\vr$ (the 25\% to 75\% quantiles) for all objects with $\dphlin<10$\,pc is similar for all four input catalogues, spanning about -30 to +30\,\kms. But the central 90\% quantile in hip2rave is -830 to +645\,\kms, compared to about -70 to +80\,\kms\ for the other three input catalogues.  These discrepancies alone do not suggest that the large Rave radial velocities are wrong.  However, of the 11 Rave objects with $\dphmean<$\,2\,pc, nine have anomalously high radial velocities, and all but one of these are listed with a low SNR from the RAVE pipeline (``SNRS'' of 1--2).
The other two objects, \object{Hip 71681}.r and \object{Hip 87784}.r, have moderate radial velocities and higher SNR (52 and 11 respectively).  This suggests problems with extracting reliable radial velocities at low SNR, because although the corresponding $\sigma(\vr)$ values are also large, they are not large enough to make these measurements consistent with the more typical radial velocities we see of below 100\,\kms.  It is important to realize that a search for close encounters preferentially includes stars with large radial velocities, anomalous or not: All other data being equal, the larger the radial velocity, the smaller the perihelion distance (as the velocity vector then points closer to the Sun).

Needless to say, I can only find encounters among the stars I have actually analysed, namely a subset of Hipparcos stars.  A few additional encounters have been found using other astrometry (e.g.\ two M dwarfs found by \citealt{2010AstL...36..816B}), and the current work could be extended using additional radial velocities.

\subsection{Correlation with impact craters}

It would be interesting to investigate whether any of the close encounters can be associated with a specific impact crater on the Earth. This would be difficult, however, for a number of reasons: (a) both this survey and the impact cratering record on the Earth are very incomplete; (b) the Galactic tide is also a major and variable Oort cloud perturber \citep{2005A&A...441..783D, 2014MNRAS.442.3653F, 2014M&PS...49....8R}; (c) there is a time lag between the perturbation occurring and a comet shower arriving in the inner solar system \citep{1998ApJ...499L.219F}. This is 500\,kyr for comets coming from $10^4$\,AU (i.e.\ half the orbital period),
but because the comets' perihelia are not reduced instantly by the perturbation, it can take a few million years until the peak of the comet shower reaches the inner solar system \citep{2002A&A...396..283D}. Given the large range of Oort cloud comet orbits, this leads to a large uncertainty in the lag. The fact that the encounter is a continuous event spread over $\delta t \sim 1\,{\rm pc}/\vph \simeq$\,30\,kyr (adopting the median of $\vphmean$ of 34\,\kms\ for encounters with $\dphmean<1$\,pc) introduces only a small time uncertainty compared to the shower lag;
(d) most impacts are probably due to asteroids rather than comets \citep{2007IAUS..236..441W, 2013AcAau..90....3Y}; 
(e) some comets may not reach the Earth, due to the shielding effect of Jupiter, for example \citep{2010IJAsB...9....1H}. 
Thus we may instead be limited to connecting the encounters with the overall time variation of the impact cratering rate \citep{2011MNRAS.416.1163B, 2011MNRAS.418.2111B}.

According to the impact approximation, the largest perturbers of the Oort cloud are the first four entries in Table~\ref{tab:periStats}, despite their relatively small masses.  The largest perturbation in the past is due to \object{gamma Microscopii} (\object{Hip 103738}.p), 3.85\,Myr ago (90\% CI of 4.16--3.56\,Myr). 
At $\dphmean$\,=\,0.83\,pc it would have appeared from the Earth to be slightly brighter than Venus at maximum.
The above arguments notwithstanding, it is alluring that there are two impact craters listed in the Earth Impact Database \cite{EID} with compatible ages (ignoring craters with only upper or lower age limits):
Roter Kamm (Namibia) -- age $3.7\pm0.3$\,Myr, diameter 2.5\,km; El'gygytgyn (Siberia) -- age $3.5\pm0.5$\,Myr, diameter 18\,km. A third, Aouelloul (Mauritania) -- age $3.0\pm0.3$\,Myr, diameter 0.4\,km -- is rather small, and two others have less well defined ages: Karla at $5\pm1$\,Myr and Bigach at $5\pm3$\,Myr. Of course, with many potential perturbers and many impact craters, some coincidences are inevitable.

\subsection{Future work}\label{sec:future}

What could improve this study? Most important would be a larger 6D phase space stellar catalogue extending to larger volumes, with better understood completeness, and reliable treatment of binarity. Starting in 2016, we can expect data releases from Gaia which ultimately promises to provide astrometry with parallax accuracies between 0.005 and 0.5\,\mas\ on around $10^9$ stars down to magnitude $G=20$ \citep{2008IAUS..248..217L, 2012Ap&SS.341...31D}.  Proper motion accuracies (in \maspyr) are similar, which is important because the best encounter candidates among more distant stars are those with very small proper motions. Gaia should also provide radial velocities on up to $10^8$ stars down to $I\simeq16$ with accuracies of 1--15\,\kms\ depending on magnitude.
This will greatly extend the reach of the analysis in this paper to a significantly larger volume, and will begin to include many more massive (and rare) stars  (see also \citealt{2011A&A...535A..86F, 2012P&SS...73..124R}). 
Yet integrating orbits over much longer paths will also make the results more sensitive to the gravitational potential adopted, although the Gaia data itself will help to improve our determination of this.
It will be important to supplement Gaia data with radial velocities on fainter stars and with more accurate radial velocities, especially for stars with radial velocities near to zero, for stars known or suspected to be binaries, and for stars with questionable current data. 

The depth and better understood completeness of Gaia will also permit a more robust and useful determination of the actual probability of encounters as a function of perihelion time and distance than was possible with Hipparcos \citep{2001A&A...379..634G}.  While it may remain difficult to connect individual stellar encounters with specific impact craters by such a study, Gaia should allow us to investigate the link between encounters and impacts in a statistical sense \citep{2014MNRAS.442.3653F}.

Another important feature of Gaia and its data processing consortium is that it will provide astrophysical parameters for all of its targets using onboard spectrophotometry and spectroscopy \citep{2013A&A...559A..74B}. Mass estimates will therefore be available for many targets, so a more systematic investigation of the perturber impulse can also be undertaken.

\begin{acknowledgements}

This research is based on data from the Hipparcos astrometry satellite and has made use of the VizieR catalogue access tool, the Simbad object database, and the cross-match service, provided by CDS, Strasbourg.  I would like to thank Floor van Leeuwen for reanalysing the Hipparcos data on the Hip 85607+85605 system.  I am grateful to Morgan Fouesneau, Fabo Feng, Nick Sleep, and an anonymous referee for helpful comments. My special thanks go to Barrie W.\ Jones (1941--2014) for his teaching, enthusiasm, and years of support.

\end{acknowledgements}

\bibliographystyle{aa}
\bibliography{stellar_encounters}

\begin{appendix}
\section{Equivalence of the data sampling to estimating the posterior probability density function}\label{sec:BayesEquivalence}

Determining the perihelion parameters $\peripar$\,=\,$(\tph, \dph, \vph)$ given the astrometric data $D$\,=\,$(\ra, \dec, \parallax, \pmra, \pmdec, \vr)$ is an inference problem. The full characterization of the solution is the probability density function (PDF) $P(\peripar | D, C, M)$ where $M$ denotes the Galaxy model, and $C$ is the covariance in $D$. In a gravitational field, seven variables are required to completely define the orbit of a star: a 3D position vector, a 3D velocity vector, and time.  One realization of this is $D$ together with the current time.  Another is the state of the star at perihelion, which is described by $\peripar$ and four additional parameters $\nupar$, two of which determine the angular position on the sky, and two which determine the direction of the velocity.  The orbital integration can be considered as just performing a coordinate transformation from $D$ (with $t=0$) to $(\peripar, \nupar)$.

The direct approach to the inference problem is to use Bayes' theorem to write the posterior PDF of the parameters as
\begin{equation}
P(\peripar, \nupar | D, C, M) \,=\, \frac{P(D | \peripar, \nupar, C, M)P(\peripar, \nupar | M)}{P(D | C, M)} \ .
\label{eqn:bayes}
\end{equation}
A common way to construct this PDF is by sampling from the unnormalized posterior (the numerator on the right-hand side) with an Markov Chain Monte Carlo (MCMC) technique.  For each proposed sample $(\peripar, \nupar)$ we would use the orbital integration backwards to generate an orbit and then to find the corresponding position in $D$-space at $t=0$.  The likelihood, $P(D | \peripar, \nupar, C, M)$, is the error model for the data (section~\ref{sec:MethodSampling}), and the prior, $P(\peripar, \nupar | M)$, is something we decide upon.  Once we have the set of samples we can generate the posterior PDF via density estimation (e.g.\ a histogram) and explicitly normalize.  We arrive at the PDF of interest by marginalizing over the ``nuisance'' parameters, $\nupar$,
\begin{equation}
P(\peripar | D, C, M) \,=\, \int P(\peripar, \nupar | D, C, M) \, d\nupar \ .
\end{equation}

The problem with this approach is that any plausible prior corresponds to a very large set of orbits which cannot be well constrained in advance. The vast majority of orbits defined by the proposed $(\peripar, \nupar)$ samples will come nowhere near to $D$, with the result that the likelihood of most samples will be essentially zero. The posterior will therefore be badly estimated unless we use a huge number of samples. In other words, this approach is too computationally inefficient to be of practical use.

The method I use in this paper (section~\ref{sec:MethodSampling}), while intuitive, is a sampling of $D$ and not of $\peripar$, so it is not immediately obvious that it yields $P(\peripar | D, C, M)$. But in fact it does when we (implicitly) adopt a uniform prior, for the following reason.  Bayes' theorem translates knowledge about the likelihood (uncertainties in the data) into knowledge about the posterior (a distribution over the parameters). This includes a prior, but if that prior shows no preference to different parameter values (i.e.\ it is uniform over the range supported by the data), then it will not change the shape of the resulting posterior.  So we only have to transform the data samples into parameter samples, and this can be done here because of the reversibility of orbits: $D$ (with $t=0$) maps one-to-one onto $(\peripar, \nupar)$.  This produces an equivalent set of samples as the MCMC approach would with a uniform prior.  The difference is that sampling the large $(\peripar, \nupar)$-space and mapping each sample to $D$-space is very inefficient, but vice versa is not, because by sampling directly in $D$-space we sample areas which give high likelihoods. It avoids sampling the enormous low likelihood region of $D$-space. We of course have to renormalize the resulting distribution, 
and this is done explicitly using a density estimate over the samples (as in Figure \ref{fig:perihistogram}), just as with MCMC.

One could argue that a uniform prior over the parameters is not appropriate, and hence that a direct parameter sampling should be used. Yet the data sampling approach could still be used, provided there is a way of mapping the parameter priors to ``priors'' on the sampled data. 
In the present case, the results will not be very sensitive to the prior given the relatively high signal-to-noise ratio of the data (that is, they are informative), so a uniform prior is justified.

\end{appendix}

\end{document}